\documentclass[twocolumn]{aastex631}
\usepackage{CJK}

\received{August 2, 2023}
\revised{October 5, 2023}
\submitjournal{the Astronomical Journal}

\shorttitle{Verifying Gaia DR3 SB1 Solutions in the Low Doppler
Semiamplitude Limit}
\shortauthors{Schmidt et al.}

\begin{document}
\begin{CJK*}{UTF8}{gbsn}

\title{Verification of Gaia DR3 Single-lined Spectroscopic Binary
Solutions With Three Transiting Low-mass Secondaries}

\author[0000-0001-8510-7365]{Stephen P.\ Schmidt}
\affiliation{William H.\ Miller III Department of Physics and Astronomy,
Johns Hopkins University, 3400 N Charles St, Baltimore, MD 21218, USA}

\author[0000-0001-5761-6779]{Kevin C.\ Schlaufman}
\affiliation{William H.\ Miller III Department of Physics and Astronomy,
Johns Hopkins University, 3400 N Charles St, Baltimore, MD 21218, USA}
\affiliation{Carnegie Institution for Science Earth \& Planets Laboratory,
5241 Broad Branch Road NW, Washington, DC 20015, USA}

\author[0000-0001-8470-1725]{Keyi Ding (丁可怿)}
\affiliation{William H.\ Miller III Department of Physics and Astronomy,
Johns Hopkins University, 3400 N Charles St, Baltimore, MD 21218, USA}

\author[0000-0003-4976-9980]{Samuel K.\ Grunblatt}
\affiliation{William H.\ Miller III Department of Physics and Astronomy,
Johns Hopkins University, 3400 N Charles St, Baltimore, MD 21218, USA}

\author[0000-0001-6416-1274]{Theron Carmichael}
\affiliation{Institute for Astronomy, University of Edinburgh, Royal
Observatory, Blackford Hill, Edinburgh, EH9 3HJ, UK}

\author[0000-0001-6637-5401]{Allyson Bieryla}
\affiliation{Center for Astrophysics \textbar \ Harvard \& Smithsonian,
60 Garden St, Cambridge, MA 02138, USA}

\author[0000-0001-8812-0565]{Joseph E.\ Rodriguez} 
\affiliation{Center for Data Intensive and Time Domain Astronomy,
Department of Physics and Astronomy, Michigan State University, East
Lansing, MI 48824, USA}

\author[0000-0002-7382-0160]{Jack Schulte}
\affiliation{Center for Data Intensive and Time Domain Astronomy,
Department of Physics and Astronomy, Michigan State University, East
Lansing, MI 48824, USA}

\author[0000-0002-0701-4005]{Noah Vowell}
\affiliation{Center for Data Intensive and Time Domain Astronomy,
Department of Physics and Astronomy, Michigan State University, East
Lansing, MI 48824, USA}

\author[0000-0002-4891-3517]{George Zhou}
\affiliation{Centre for Astrophysics, University of Southern Queensland,
West Street, Toowoomba, QLD 4350, Australia}

\author[0000-0002-8964-8377]{Samuel N.\ Quinn}
\affiliation{Center for Astrophysics \textbar \ Harvard \& Smithsonian,
60 Garden St, Cambridge, MA 02138, USA}

\author[0000-0001-7961-3907]{Samuel W.\ Yee}
\affiliation{Department of Astrophysical Sciences, Princeton University,
4 Ivy Lane, Princeton, NJ 08544, USA}

\author[0000-0002-4265-047X]{Joshua N.\ Winn}
\affiliation{Department of Astrophysical Sciences, Princeton University,
4 Ivy Lane, Princeton, NJ 08544, USA}

\author[0000-0001-8732-6166]{Joel D.\ Hartman}
\affiliation{Department of Astrophysical Sciences, Princeton University,
4 Ivy Lane, Princeton, NJ 08544, USA}

\author[0000-0001-9911-7388]{David W.\ Latham}
\affiliation{Center for Astrophysics \textbar \ Harvard \& Smithsonian,
60 Garden St, Cambridge, MA 02138, USA}

\author[0000-0003-1963-9616]{Douglas A.\ Caldwell}
\affiliation{SETI Institute, Moffett Field, Mountain View, CA, 94035, USA}
\affiliation{NASA Ames Research Center, Moffett Field, CA 94035, US}

\author[0000-0002-9113-7162]{M.\ M.\ Fausnaugh}
\affiliation{Department of Physics and Kavli Institute for Astrophysics
and Space Research, Massachusetts Institute of Technology, Cambridge,
MA 02139, USA}
\affiliation{Department of Physics \& Astronomy, Texas Tech University,
Lubbock, TX, 79410-1051, USA}

\author[0000-0002-3385-8391]{Christina Hedges}
\affiliation{University of Maryland, Baltimore County, 1000 Hilltop Cir,
Baltimore, MD 21250, USA}
\affiliation{NASA Goddard Space Flight Center, 8800 Greenbelt Rd,
Greenbelt, MD 20771, USA}

\author[0000-0002-4715-9460]{Jon M.\ Jenkins}
\affiliation{NASA Ames Research Center, Moffett Field, CA 94035, US}

\author[0000-0002-4047-4724]{Hugh P.\ Osborn}
\affiliation{Department of Physics and Kavli Institute for Astrophysics
and Space Research, Massachusetts Institute of Technology, Cambridge,
MA 02139, USA}
\affiliation{NCCR/Planet-S, Universit\"{a}t Bern, Gesellschaftsstrasse 6,
3012 Bern, Switzerland}

\author[0000-0002-6892-6948]{S.\ Seager}
\affiliation{Department of Physics and Kavli Institute for Astrophysics
and Space Research, Massachusetts Institute of Technology, Cambridge,
MA 02139, USA}
\affiliation{Department of Earth, Atmospheric and Planetary Sciences,
Massachusetts Institute of Technology, Cambridge, MA 02139, USA}
\affiliation{Department of Aeronautics and Astronautics, MIT, 77
Massachusetts Avenue, Cambridge, MA 02139, USA}

\correspondingauthor{Stephen Schmidt}
\email{sschmi42@jh.edu}

\begin{abstract}

\noindent
While secondary mass inferences based on single-lined spectroscopic binary
(SB1) solutions are subject to $\sin{i}$ degeneracies, this degeneracy can
be lifted through the observations of eclipses.  We combine the subset of
Gaia Data Release (DR) 3 SB1 solutions consistent with brown dwarf-mass
secondaries with the Transiting Exoplanet Survey Satellite (TESS)
Object of Interest (TOI) list to identify three candidate transiting
brown dwarf systems.  Ground-based precision radial velocity follow-up
observations confirm that TOI-2533.01 is a transiting brown dwarf with
$M=72^{+3}_{-3}~M_{\text{Jup}}= 0.069^{+0.003}_{-0.003}~M_\odot$ orbiting
TYC 2010-124-1 and that TOI-5427.01 is a transiting very low-mass star
with $M=93^{+2}_{-2}~M_{\text{Jup}}=0.088^{+0.002}_{-0.002}~M_\odot$
orbiting UCAC4 515-012898.  We validate TOI-1712.01 as a very low-mass
star with $M=82^{+7}_{-7}~M_{\text{Jup}}=0.079^{+0.007}_{-0.007}~M_\odot$
transiting the primary in the hierarchical triple system BD+45 1593.
Even after accounting for third light, TOI-1712.01 has radius nearly
a factor of two larger than predicted for isolated stars with similar
properties.  We propose that the intense instellation experienced
by TOI-1712.01 diminishes the temperature gradient near its surface,
suppresses convection, and leads to its inflated radius.  Our analyses
verify Gaia DR3 SB1 solutions in the low Doppler semiamplitude limit,
thereby providing the foundation for future joint analyses of Gaia radial
velocities and Kepler, K2, TESS, and PLAnetary Transits and Oscillations
(PLATO) light curves for the characterization of transiting massive
brown dwarfs and very low-mass stars.

\end{abstract}


\section{Introduction}\label{sec:intro}

While isochrones can be used to infer the masses and radii of isolated
low-mass stars and brown dwarfs, the theoretical models underlying 
those isochrones have difficulties reproducing the measured masses
and radii of such objects \citep{Kraus2011,Birkby2012}.  On the other
hand, the minimum masses of the secondary stars and brown dwarfs in
single-lined spectroscopic binary systems (SB1s) can be measured
relative to the masses of the primaries in such systems.  If the 
photospheres of the primaries in such systems are much, much hotter than
the temperatures of the secondaries' ``night'' sides and the secondaries
have significantly smaller radii (both of which apply for the lowest-mass
stellar secondaries), then the eclipses of such systems can be treated
like the transits and occultations of exoplanet systems.  In that case,
the depth of an eclipse of the primary by the secondary can be used to
measure the radius of the low-mass secondary relative to the radius of
the primary.  The eclipsing nature of such a system lifts the $\sin{i}$
degeneracy of an SB1 solution as well.  If the primary is a Sun-like
star, then isochrone-inferred masses and radii can be both accurate
and precise.  Consequently, a homogeneous analysis of a large sample of
transiting single-lined spectroscopic binaries would be a powerful way
to infer accurate and precise masses and radii for a large population
of low-mass stars and brown dwarfs.

Gaia Data Release (DR) 3's non-single stars catalog
\citep{GaiaDR3Summary,GaiaDR3Validation,DR3_velocities,2022gdr3.reptE...7P}
presented a list of 181,327 single-lined spectroscopic binary
solutions.  This collection of SB1s is almost two orders of
magnitudes larger than previous catalogs \citep[e.g.,][]{SB9}.
Of these solutions, 32,455 have Sun-like primaries \citep[F5V
through K5V according to][]{2013ApJS..208....9P} based on
absolute Gaia $G$-band magnitudes and effective temperatures
$T_{\text{eff}}$ from Gaia DR3's astrophysical parameters data product
\citep{GaiaDR3astrophysparams,GaiaDR3astrophysparams2}.  Of these, 2,490
secondaries have minimum masses in the brown dwarf regime defined by an
upper mass cutoff of 78.5 Jupiter masses \citep{2023A&A...671A.119C}.
Because of Gaia's limited time baseline and Doppler precision,
these systems usually have short periods and relatively high transit
probabilities.  These SB1 solutions have been validated down to the brown
dwarf mass regime \citep{GaiaDR3stellarmultiplicity}.  The infrequent
sampling of light curves based on Gaia photometry leaves them insensitive
to all but the shortest-period eclipsing systems.

NASA's Transiting Exoplanet Survey Satellite (TESS) mission has observed
most of the sky in search of exoplanets and other transiting objects
\citep{2014SPIE.9143E..20R}.  TESS light curves have been exhaustively
searched for candidate transiting planets \citep{TOICatalog}.  Now in
its Second Extended Mission, TESS data has led to the discovery
of thousands of eclipsing binaries \citep{2022ApJS..258...16P}
and over a dozen transiting very low-mass stars and brown dwarfs
\citep[e.g.,][]{populatingthebrowndwarfandstellarboundary,2022MNRAS.514.4944C,
threebrowndwarfs,2023MNRAS.523.6162L,2023AJ....165..268V}.  The population
of TESS-discovered eclipsing or transiting systems is biased towards
periods shorter than about 10 days due to TESS's observing strategy and
the smaller transit probabilities of systems with longer periods.

The combination of Gaia SB1 solutions and TESS light curves would be
a powerful resource for the accurate and precise characterization of
low-mass stars and brown dwarfs.  In this article, we seek out systems
with Gaia SB1 solutions consistent with brown dwarf-mass secondaries that
have also been observed by TESS to transit their primaries.  We identify
three such systems for which the SB1 period is well matched by the
TESS-inferred period: BD+45 1593 (TOI-1712), TYC 2010-124-1 (TOI-2533),
and UCAC4 515-012898 (TOI-5427).  Our ground-based follow-up analysis
subsequently reveals that BD+45 1593 Ab (TOI-1712.01) is a very low-mass
star transiting the primary in a hierarchical triple, TYC 2010-124-1
b (TOI-2533.01) is a transiting brown dwarf, and UCAC4 515-012898 b
(TOI-5427.01) is a transiting very low-mass star.  We describe the
initial identification of these three objects as potential brown dwarfs
in Section \ref{sec:targets}.  We describe our follow-up observations for
the systems and our subsequent analysis to confirm or validate the three
objects in Section \ref{sec:followup}.  We discuss the implications of
these discoveries in Section \ref{sec:disc} and summarize our conclusions
in Section \ref{sec:summary}.

\section{Candidate Brown Dwarf Identification}\label{sec:targets}

We first cross-match Gaia Data Release 3's non-single stars product
with the TESS Object of Interest (TOI) catalog and the TESS eclipsing
binary catalog \citep{2022ApJS..258...16P}.  We next focus on systems
where the orbital period from the Gaia SB1 solution is within 0.3\%
of the period found by TESS.  We then use the binary mass function
\begin{equation}\label{binarymassfunction}
   f(M) = \frac{M_{2}^3 \sin^3{i}}{\left(M_{1} + M_{2}\right)^2} = \frac{P K_1^3}{2 \pi G} \left(1 - e^2\right)^{3/2},
\end{equation}
in the $M_2 \ll M_1$ limit to calculate a preliminary minimum mass for
each secondary:
\begin{equation} \label{eq:radialvelocity}
   \frac{M_{2} \sin{i}}{M_{\text{Jup}}} = \left(\frac{K_{1}}{28.4329~\text{m s$^{-1}$}}\right)\sqrt{1 - e^2} \left(M_{1}\right)^{2/3} P^{1/3}.
\end{equation}
The inputs to Equation (\ref{eq:radialvelocity}) are a primary mass
$M_1$ in Solar masses from Gaia's Final Luminosity Age Mass Estimator
(FLAME) and a system's Gaia DR3 SB1 solution orbital parameters Doppler
semiamplitude $K_1$ in meters per second and period $P$ in years.
In cases in which a FLAME-derived mass is unavailable, we use the
primary mass from the TESS Follow-up Observing Program (TFOP) instead.
For systems in which secondary mass is comparable to that of a brown
dwarf, we refine our preliminary mass estimates by calculating more
precise stellar parameters for the primary.

\begin{figure*}
    \centering
    \includegraphics[width=\linewidth]{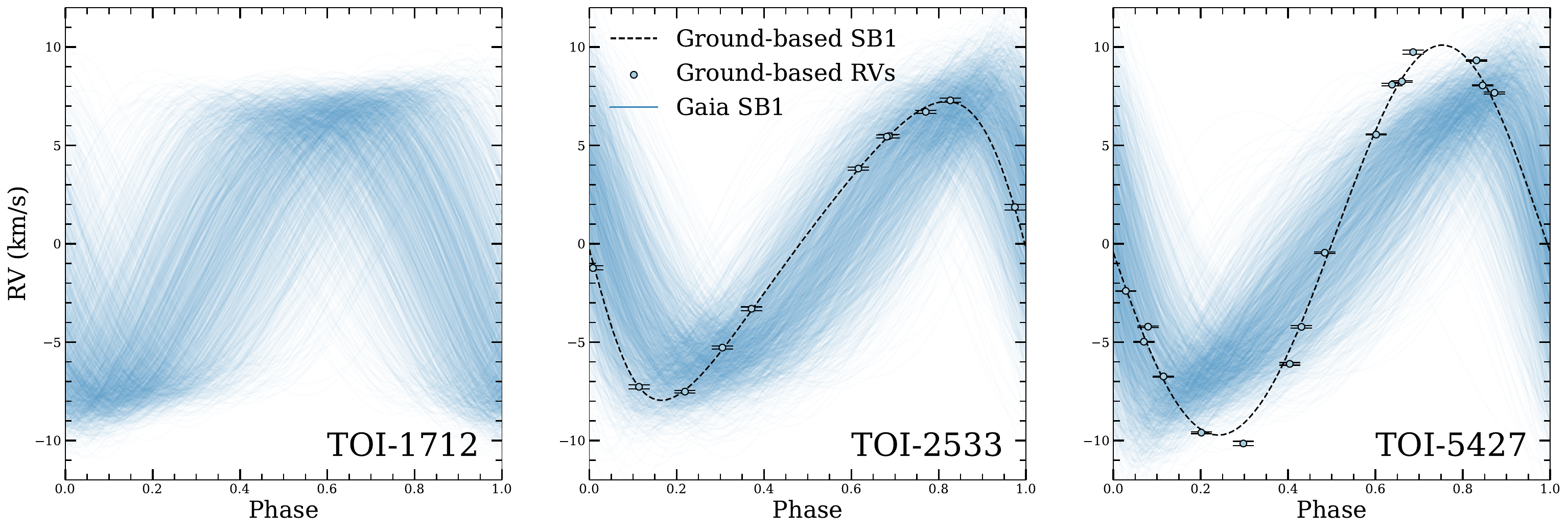}
    \caption{Doppler curves for BD+45 1593, TYC 2010-124-1, and UCAC4
    515-012898 based on their Gaia DR3 and ground-based SB1 solutions.
    We plot 2000 partially transparent blue lines to represent individual
    solutions sampled from normal distributions of the Gaia Doppler
    semiamplitude, eccentricity, and argument of periastron, with the
    standard deviation of each distribution representing the reported
    error on its respective parameter.  Gaia's reported uncertainties
    on each parameter are derived from the variance-covariance matrix of
    the solution, but the matrix itself is not reported; we therefore do
    not correct for it in this plot.  We plot the individual precision
    ground-based radial velocity measurements and their uncertainties
    as light blue circles with black borders and error bars.  We plot
    the Doppler curves inferred from these data as dashed black lines.
    The agreement between the Gaia and precision ground-based SB1
    solutions demonstrates the fidelity of the Gaia SB1 solutions at
    low Doppler semiamplitude.}
    \label{fig:sb1}
\end{figure*}

This procedure yields three systems with candidate brown dwarfs:
BD+45 1593 (Gaia DR3 917823348536224768/TOI-1712), TYC 2010-124-1
(Gaia DR3 1259922196651254272/TOI-2533), and UCAC4 515-012898 (Gaia DR3
3341401062125577088/TOI-5427).  We illustrate these systems in Figure
\ref{fig:sb1} and describe them in detail below.  These SB1 solutions
are based on Gaia $R \approx 11500$ Radial Velocity Spectrometer (RVS)
spectra that cover the wavelength range $846 \text{~nm} \lesssim \lambda
\lesssim 870 \text{~nm}$ \citep{DR3_velocities}.  The SB1 solutions
for BD+45 1593 used 18 ``good'' radial velocity (RV) measurements, TYC
2010-124-1 used 47 ``good'' RV measurements, and UCAC4 515-012898 used 25
``good'' RV measurements.  We refer to these systems by their TESS Object
of Interest designations TOI-1712, TOI-2533, and TOI-5427 in the context
of analyses based on data sourced solely from TESS or TFOP.  In all
other context, we refer to these systems by their Simbad designations
BD+45 1593, TYC 2010-124-1, and UCAC4 515-012898.

\begin{deluxetable}{ccc}
\centering
\tabletypesize{\scriptsize}
\tablewidth{0pt}
\tablecaption{TESS Observations of the Three Objects \label{tab:TESS}}
\tablehead{\colhead{Sector} &
\colhead{Corresponding Time Period} &
\colhead{Cadence (Minutes)}}
\startdata
\multicolumn{3}{c}{TOI-1712 (TIC 67926921)}\\
\hline
20 & December 24, 2019 - January 20, 2020 & 30\\
47 & December 30, 2021 - January 28, 2022 & 2, 10\\
\hline
\multicolumn{3}{c}{TOI-2533 (TIC 418012030)}\\
\hline
23 & March 19, 2020 - April 15, 2020 & 30\\
\hline
\multicolumn{3}{c}{TOI-5427 (TIC 52420398)}\\
\hline
6 & December 12, 2018 - January 6, 2019 & 30\\
43 & September 16, 2021 - October 12, 2021 & 10\\
44 & October 12, 2021 - November 6, 2021 & 10\\
45 & November 6, 2021 - December 2, 2021 & 10\\
\enddata
\end{deluxetable}

TOI-1712 and TOI-5427 were discovered by the TESS Science Office,
while TOI-2533 was first discovered by Rafael Brahm.  TOI-2533 then
became a Community TESS Object of Interest (CTOI), and later
was adopted as a TOI by the TESS Science Office.  TOI-1712 and
TOI-2533 were identified by the TESS Quick-Look Pipeline \citep[QLP
-][]{2020RNAAS...4..204H,2020RNAAS...4..206H}, while TOI-5427 was
identified by the TESS QLP Faint-star Search \citep{2022ApJS..259...33K}.
We give in Table \ref{tab:TESS} the details of the TESS observations of
the three objects.  For stars like the primaries in these systems that
were not allocated postage stamps, TESS collected observations with a
30 minute cadence during its Prime Mission and a 10/2 minute cadence
during its First/Second Extended Mission.

\begin{figure*}
    \centering
    \includegraphics[width=\linewidth]{ 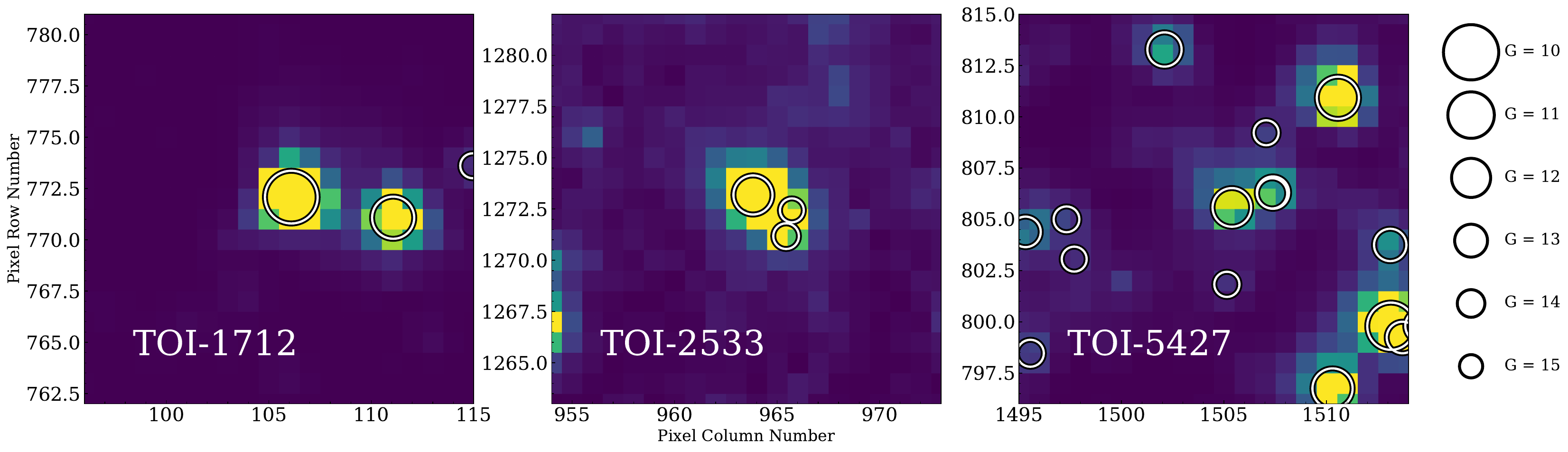}
    \caption{TESS Full-Frame Image (FFI) approximately $2048 \times
    2048$ pixel postage stamps for TOI-1712, TOI-2533, and TOI-5427.
    TESS collected the FFI for TOI-1712 during Sector 20, the FFI for
    TOI-2533 during Sector 23, and the FFI for TOI-5427 during Sector 6.
    We center each postage stamp on the target and mark each $G < 15$
    Gaia DR3 source with a white circle of radius inversely proportional
    to the Gaia $G$ magnitude.  These FFIs show that blending is not a
    concern for our light curves of the objects.}
    \label{fig:ffi}
\end{figure*}

Because of TESS's large pixels, third light can make it difficult to
interpret transit depths directly as secondary-to-primary radius ratios.
To investigate the possibility of third light in these systems, we do two
things.  We first confirm that the TESS contamination ratios\footnote{TESS
contamination ratios are given as the column \texttt{contratio} in the
TESS Input Catalog \citep{TIC2019}.} are small for each system: 0.00054
for TOI-1712, 0.039224 for TOI-2533, and 0.263055 for TOI-5427.  We next
use the high spatial resolution Gaia DR3 source catalog and the plotting
function within the \texttt{giants} pipeline \citep{2022AJ....163...53S,
2022AJ....163..120G} to show that there are no bright sources within the
TESS \texttt{lightkurve}-generated \citep{2018ascl.soft12013L} Full Frame
Image (FFI) PSF of any of these three systems (Figure \ref{fig:ffi}).

\begin{figure*}
    \centering
    \includegraphics[width=.88 \linewidth]{ 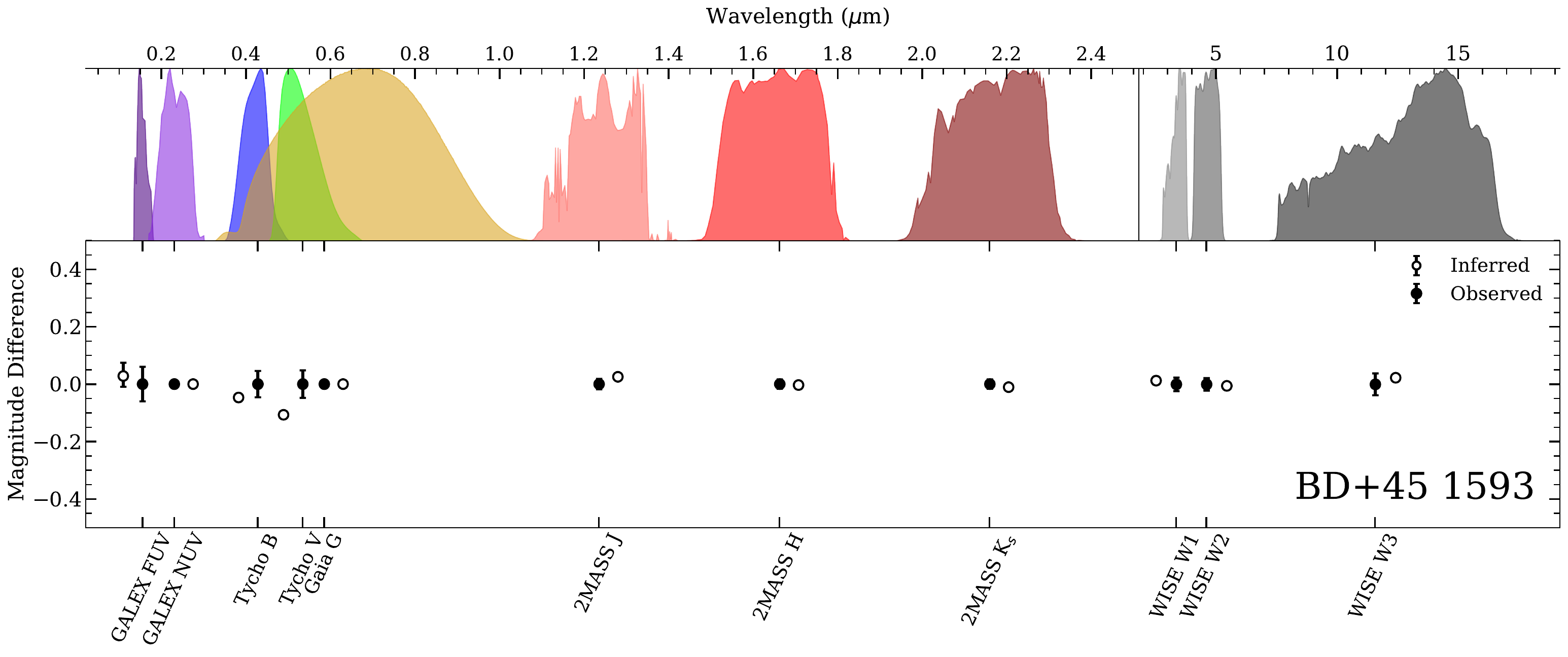}
    \includegraphics[width=.88 \linewidth]{ 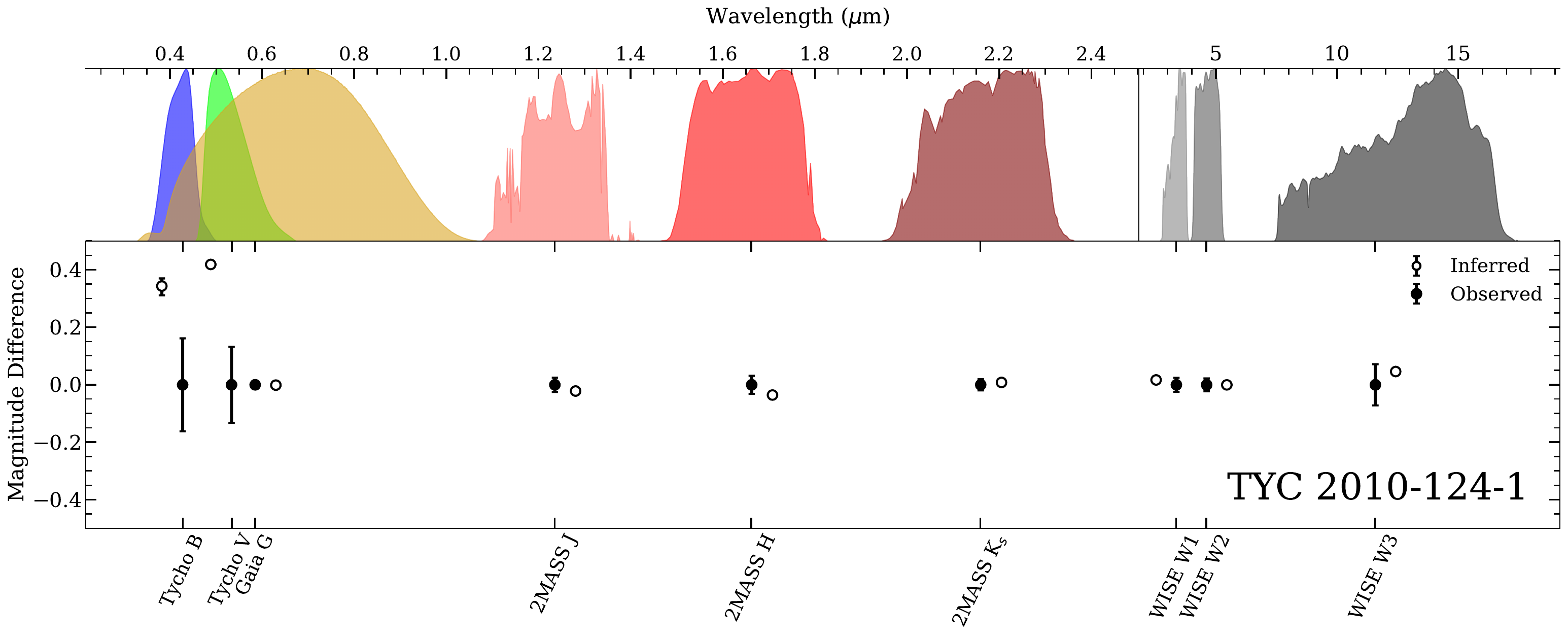}
    \includegraphics[width=.88 \linewidth]{ 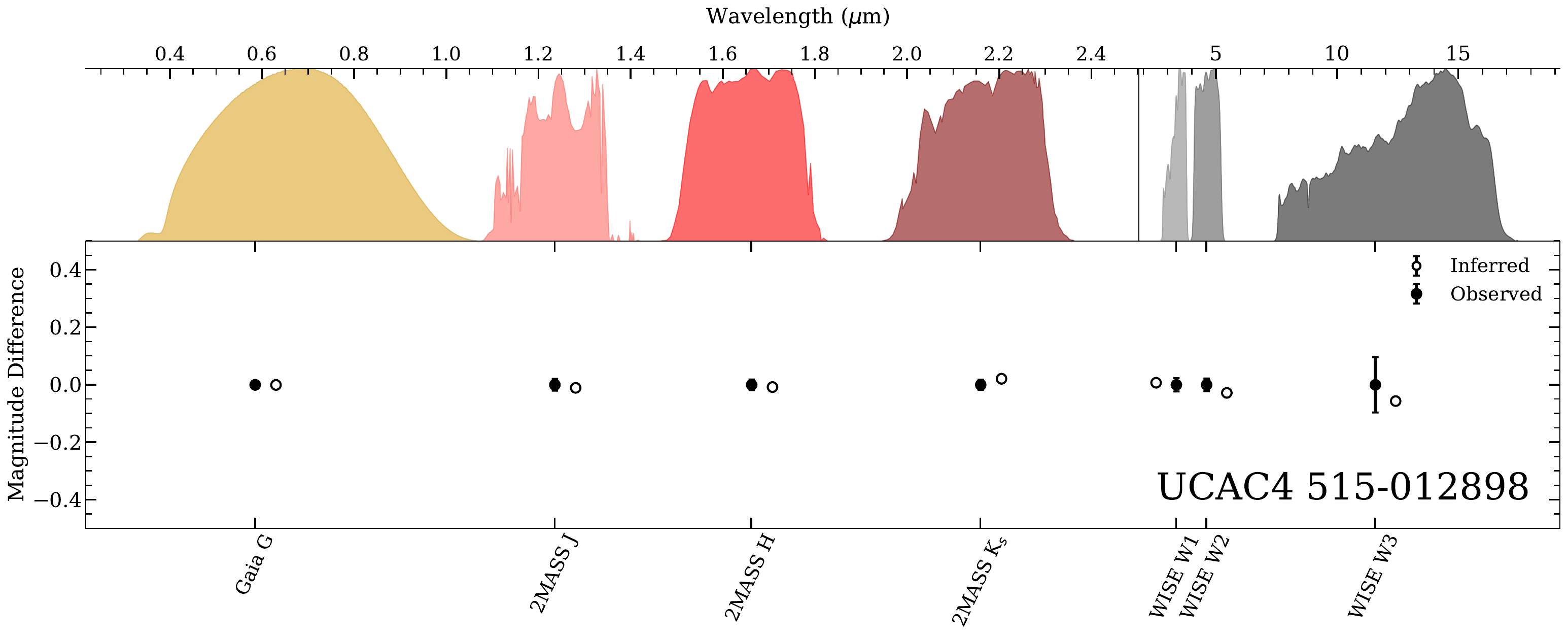}
    \caption{Comparison between the observed magnitudes for the three
    systems and the isochrone-derived magnitudes of the combined binary
    system (in the case of BD+45 1593) or the primaries (in the case
    of TYC 2010-124-1 and UCAC4 515-012898).  For each filter, we plot
    filled circles to represent the observed magnitudes and unfilled
    circles to represent the isochrone-inferred magnitudes.  We plot the
    relative transmission for each filter obtained from the SVO Filter
    Profile Service \citep{2012ivoa.rept.1015R,2020sea..confE.182R} above
    its photometric pair.  There is a discontinuity on the wavelength
    axis at the top of each plot between the 2MASS magnitudes and
    AllWISE magnitudes.  The Tycho-2 magnitudes for TYC 2010-124-1
    are close to the catalog's limit of reliability \citep{Hog2000}.
    The $B_{T}$ magnitude difference corresponds to $2.1~\sigma$, which is
    unsurprising given the number of comparisons in our sample.  However,
    the $V_{T}$ magnitude difference corresponds to $3.2~\sigma$, which
    we attribute to underestimated random uncertainties or systematic
    uncertainties at the faint end of the Tycho-2 photometric calibration.
    The consistency and overall small offsets between the observed
    and isochrone-inferred magnitudes demonstrates that our method of
    deriving the host stars' parameters is robust, especially when UV
    photometry is available.}
    \label{fig:magcomp}
\end{figure*}

We derive the fundamental and photospheric stellar parameters of BD+45
1593, TYC 2010-124-1, and UCAC4 515-012898 using the \texttt{isochrones}
\citep{mor15} package to execute with \texttt{MultiNest}
\citep{fer08,fer09,fer19} a simultaneous Bayesian fit of the MIST
isochrone grid \citep{pax11,pax13,pax18,pax19,jer23,dot16,cho16} to
a curated collection of data for each star.  The stellar parameters
derived using this approach have been shown by Ding et al. (2024,
in prep) to reproduce the asteroseismology-inferred masses and
radii of the Kepler LEGACY Sample \citep{2017ApJ...835..173S}.
We fit (1) a zero point-corrected Gaia DR3 parallax
\citep{gai21,fab21,lin21a,lin21b,row21,tor21}, (2) Galaxy
Evolution Explorer (GALEX) GUVcat\_AIS NUV and FUV \citep{bia17},
Tycho-2 $B_{T}$ and $V_{T}$ \citep{Hog2000}, Gaia DR2 $G$
\citep{gai16,gai18,are18,eva18,rie18}, Two-micron All-sky Survey (2MASS)
$JHK_{\text{s}}$ \citep{skr06}, and/or Wide-field Infrared Survey
Explorer AllWISE $W1W2W3$ photometry \citep{wri10,allwise}, and (3) an
estimated extinction value $A_V$ based on three-dimensional reddening
maps \citep{Reddening_1,Reddening_2}.  GALEX photometry is unavailable
for TYC 2010-124-1 and UCAC4 515-012898, and Tycho-2 photometry is
unavailable for UCAC4 515-012898.  We use a uniform extinction prior
bounded by $A_V \pm 5\sigma_{A_V}$ and a distance prior proportional to
volume bounded by the \citet{2021AJ....161..147B} lower and upper bounds
on geometric distances.  We report the photospheric stellar parameters
in Table \ref{tab:prop} and compare the isochrone-derived magnitudes to
each star's observed magnitudes in Figure \ref{fig:magcomp}.

As we will show in the next section, there is evidence that BD+45 1593
is a double-lined spectroscopic binary.  To determine the properties
of both luminous stars in the system, we use an \texttt{isochrones}
binary star model that requires both the primary and secondary to have
the same age, composition, and distance.  We use the full set of inputs
mentioned previously.  In Section \ref{sec:lightcurves}, we will use the
TESS $T$-band flux ratio so inferred to show that the dilution of the
transit in the BD+45 1593 system is small.  The modeled photometry that
results from the \texttt{isochrones} is consistent with the observed
photometry, and this consistency supports the accuracy of the binary
star model output (Figure \ref{fig:magcomp}).

Our isochrone analysis gives us several reasons to believe that the
low-mass object in the BD+45 1593 system orbits the most massive component
BD+45 1593 A.  We infer $T$-band magnitudes $9.95^{+0.05}_{-0.03}$ for
the primary BD+45 1593 A and $13.90^{+2.47}_{-1.12}$ for the secondary
BD+45 1593 B.  Given that the posterior-derived Gaia DR2 $G$ magnitudes for
the two stars are $10.30^{+0.05}_{-0.02}$ and $14.46^{+2.81}_{-1.25}$,
we argue that this flux difference would have left Gaia insensitive
to the possible RV variation of BD+45 1593 B.  As we will show, the
second set of lines in Tillinghast Reflector Echelle Spectrograph
reconnaissance spectra of BD+45 1593 do not move on the time scale
of months.  On the other hand, the agreement of the Gaia DR3 SB1 period
and the transit-inferred TOI period suggests that the object of interest
in the system orbits BD+45 1593 A rather than BD+45 1593 B.  Therefore,
we refer to the primary of this system as BD+45 1593 Aa and report its
parameters in Table \ref{tab:prop}.

We find that BD+45 1593 Aa is a subgiant with $M_\ast \approx
1.63^{+0.01}_{-0.02}~M_{\odot}$.  It has been shown that the rotation
period of a subgiant is proportional to its mass and therefore age
\citep{2013ApJ...776...67V}.  Assuming an aligned orbit, our inferred
mass for BD+45 1593 Aa is large enough that Gaia should be able to detect
the rotational broadening of its absorption lines.  Indeed, the Gaia
DR3 \texttt{vbroad} parameter for BD+45 1593 is $30 \pm 6$ km s$^{-1}$.
We assume that all of the line broadening measured by the Gaia DR3
\texttt{vbroad} parameter is a result of rotation and use it to calculate
the rotation period of BD+45 1593 Aa.  We divide the isochrone-inferred
circumference of BD+45 1593 Aa by its Gaia DR3 \texttt{vbroad} parameter,
resulting in a rotation period $P_{\text{rot}} \approx 5$ days.  It is
possible that angular momentum exchange between the system's revolution
and the rotation of BD+45 1593 Aa would lead to its faster rotation,
though in that case one would expect that the system's orbital period $P
= 3.566276 \pm 0.0000044$ days and the rotation period of the primary
$P_{\text{rot}} \approx 5$ days would be synchronized.  The agreement
between our inferred rotation period $P_{\text{rot}} \approx 5$ and
the \citet{2013ApJ...776...67V}-predicted slow-launch rotation period
$P_{\text{slow}} \approx 6$ days for a star of the mass, composition,
and temperature of BD+45 1593 Aa supports our inference that BD+45 1593
Aa is massive.

\section{Follow-up Observations and Analysis} \label{sec:followup}

\subsection{Light Curve Analysis} \label{sec:lightcurves}

The Box-fitting Least Squares \citep[BLS -][]{BLS} algorithm-based
transit depths reported in the TOI catalog are only indicative of
true transit depths.  We therefore reanalyze the TESS light curves of
TOI-1712, TOI-2533, and TOI-5427 with the \texttt{juliet} Python package
\citep{2019MNRAS.490.2262E} to obtain more accurate and precise radius
ratio posteriors for these three systems.  We retrieve the light curve
data products originally reduced by the TESS Science Processing Operations
Center \citep[SPOC -][]{2016SPIE.9913E..3EJ} for the three systems from
the Mikulski Archive for Space Telescopes (MAST).  For each system,
we use the available 30-minute cadence light curves produced from TESS
Full-Frame Images by the TESS-SPOC High Level Science Products (HLSP)
project \citep{2020RNAAS...4..201C}.  For TOI-1712, we additionally
use the 2-minute cadence Presearch Data Conditioned \citep[PDC-SAP
-][]{Stumpe2012,Stumpe2014,Smith2012} light curve from Sector 47 produced
by the SPOC.  Likewise, for TOI-5427 we additionally use 10-minute cadence
light curves from Sectors 43, 44, and 45 also produced from TESS FFIs
by the TESS-SPOC HLSP project.

\begin{figure*}
    \centering
    \includegraphics[width=\linewidth]{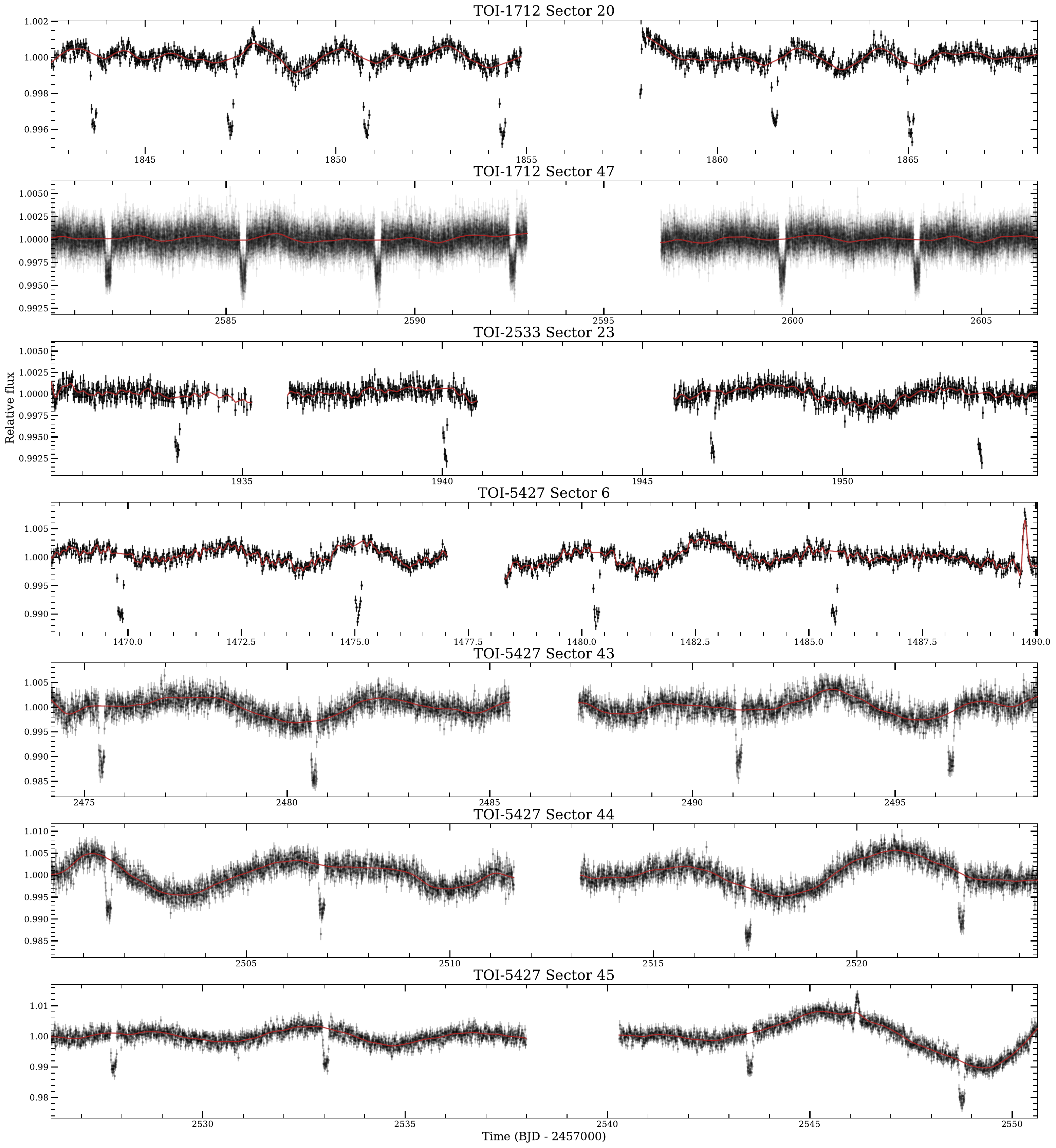}
    \caption{TESS Science Processing Operations Center (SPOC)-reduced
    light curves for TOI-1712, TOI-2533, and TOI-5427.  In addition to the
    30-minute cadence data for each of the objects, we include two-minute
    cadence data for TOI-1712 from TESS Sector 47 and ten-minute cadence
    data for TOI-5427 from Sectors 43, 44, and 45.  We plot in red on
    top of the data the \texttt{juliet}-derived Gaussian process fits
    to the light curves.  We next use these fits to detrend the light
    curves before subsequently fitting the transits.}
    \label{fig:gp}
\end{figure*}

We mask the transits in each light curve based on the ephemeris from
the TOI catalog and use the Gaussian Process (GP) modeling tool in
\texttt{juliet} to model out the stellar variability.  We use a simple
approximate Matern kernel for this process.  For the dilution factor we
use a fixed prior at 1, for the mean out-of-transit flux we use a Gaussian
prior centered at 0 with a standard deviation of 0.1, for the jitter
term $\sigma_w$ and amplitude of the GP $\sigma_{GP}$ we use logarithmic
uniform priors between $10^{-6}$ and $10^6$, and for the Matern time-scale
of the GP $\rho_{GP}$ we use a logarithmic uniform prior between $10^{-3}$
and $10^3$.  We plot the systems' original TESS light curves along with
the Gaussian process stellar variability fit in Figure \ref{fig:gp}.

\begin{figure*}
    \centering
    \includegraphics[width=\linewidth]{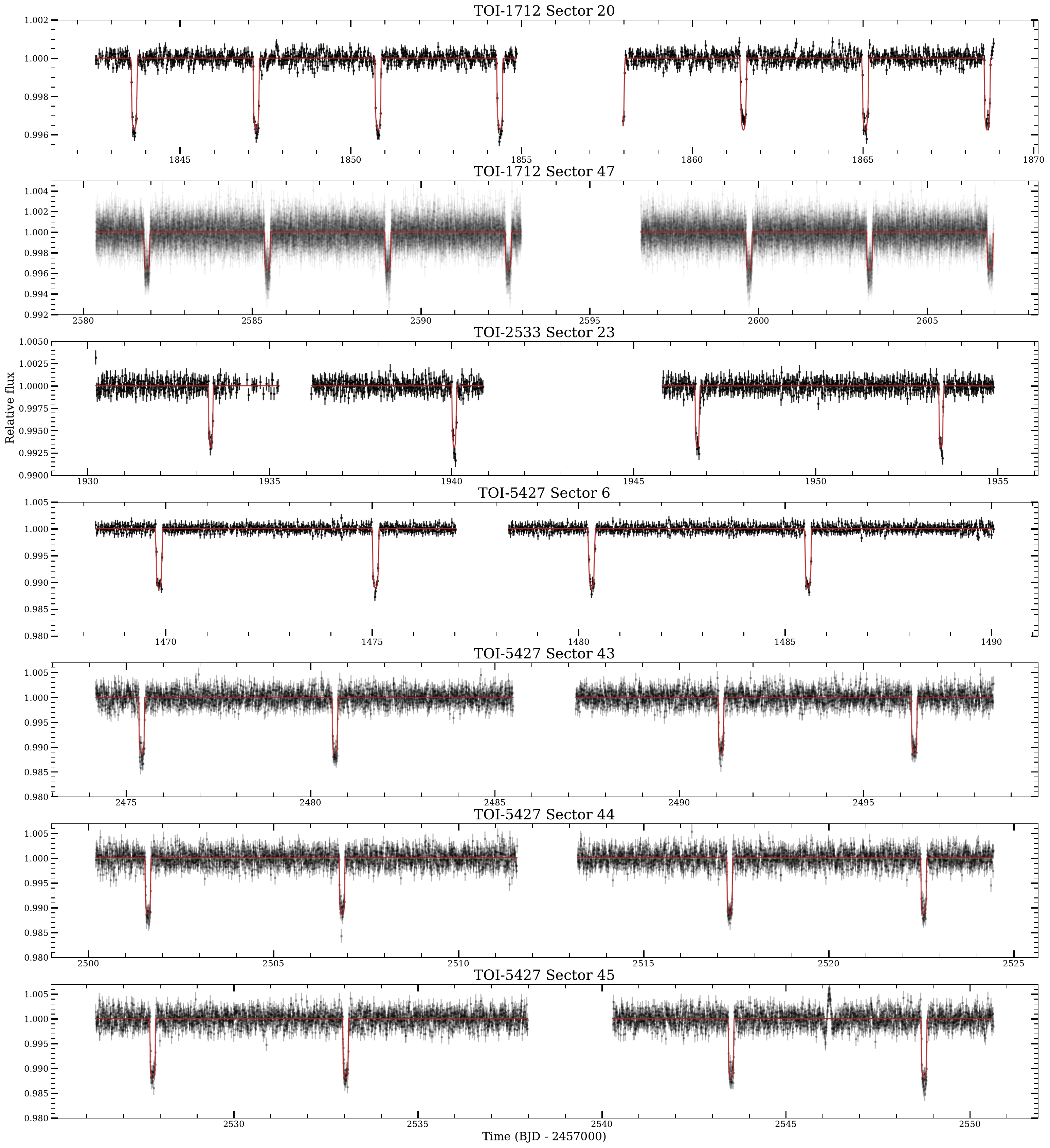}
    \caption{Detrended light curves for TOI-1712, TOI-2533, and TOI-5427.
    In addition to the 30-minute cadence data for each of the objects,
    we include two-minute cadence data for TOI-1712 from TESS Sector
    47 and ten-minute cadence data for TOI-5427 from Sectors 43,
    44, and 45.  We overlay the data with red lines representing our
    \texttt{juliet}-derived transit fits.  These fits demonstrates that
    the periods and transit shape parameters reported in the TOI catalog
    accurately reflect the data and are sufficient for comparison to
    external catalogs.}
    \label{fig:lc}
\end{figure*}

\begin{figure*}
    \centering
    \includegraphics[width=\linewidth]{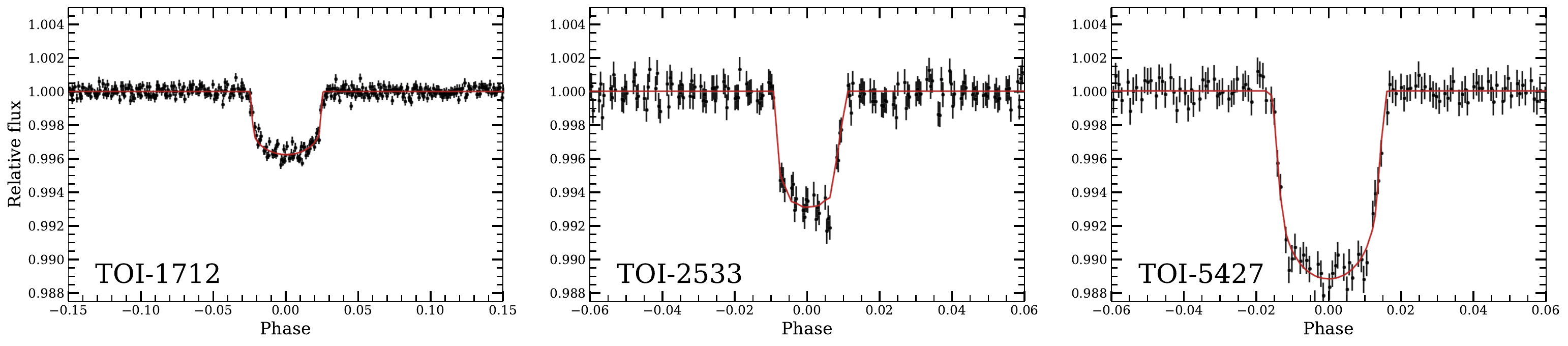}
    \caption{Phased light curves for the 30-minute cadence data
    of TOI-1712, TOI-2533, and TOI-5427.  We overlay the data with
    red lines representing the \texttt{juliet}-derived transit fits.
    The relative flux axes are consistent across the three objects to
    illustrate their differing transit depths.}
    \label{fig:phased}
\end{figure*}

After removing stellar variability, for each light curve we unmask the
transits and use the TOI ephemeris as a prior on the \texttt{juliet} fit
to extract inclination and radius ratio posteriors.  We use similar priors
as before, except we use the \citet{Kipping2013} quadratic limb-darkening
parameterization with \texttt{q1} and \texttt{q2} $\mathcal{U}(0,1)$
priors appropriate for TESS.  We plot in Figure \ref{fig:lc} the detrended
light curves for the systems with their transit fits, while in Figure
\ref{fig:phased} we plot the phase-folded 30 minute-cadence light curves
and transit fit for each system.  We report the inclinations and radius
ratios in Table \ref{tab:prop}.

\subsection{Doppler Analysis}\label{sec:rvanalysis}

We present in Table \ref{tab:rvs} 27 ground-based precision radial
velocity measurements we obtained for TOI-2533 and TOI-5427.  We collected
11 spectra for TOI-2533 between March 5, 2021 and April 8, 2021 using the
Tillinghast Reflector Echelle Spectrograph \citep[TRES -][]{gaborthesis}
on the 1.5-m Tillinghast Reflector telescope, situated on Mt.\ Hopkins
in Arizona, USA.  TRES has $R \approx 44000$ over a wavelength range
approximately 3900 \AA~to 9100~\AA.  The exposure times ranged from 2200
s to 3600 s with signal-to-noise per resolution element (S/N) between
23 and 35.  We visually reviewed each order to remove cosmic rays.
Depending on the S/N, we used 10-14 echelle orders within this wavelength
range for each spectrum to measure each order's relative radial velocity.
We used the spectrum with highest S/N per resolution element as the
reference value for these relative radial velocities.

We also collected three spectra for TOI-1712 between March 7, 2020
and February 3, 2023 using TRES.  The exposure times ranged from 600
s to 2160 s and with S/N between 35 and 67.  These spectra revealed a
cross-correlation function indicative of a double-lined spectroscopic
binary.  The evolution of the cross-correlation function suggests that
(1) the primary shows some orbital motion as seen by Gaia and (2) the
secondary appears stationary.  Combined with the Gaia DR3 SB1 solution,
this observation suggests that BD+45 1593 (TOI-1712) is a hierarchical
triple system with an unresolved secondary diluting the transit observed
by TESS.

We collected 11 spectra of TOI-5427 between September 25, 2022 and
October 7, 2022 using the CTIO High Resolution Spectrometer \citep[CHIRON
-][]{CHIRON} high-resolution spectrograph, installed in the 1.5-m
telescope at the Cerro Tololo International Observatory.  We used an
image slicer that produces spectra with $R \approx 80000$.  The exposure
times ranged between 3000 s and 3650 s with S/N between 47 and 82 at
550 nm.  We calibrated each science spectrum with a thorium-argon (ThAr)
spectrum taken immediately beforehand to account for instrumental spectral
drift and to compute a new wavelength solution automatically from this
calibration with the CHIRON pipeline \citep{CHIRONPipeline}.  To extract
the radial velocities from these CHIRON observations, we used a least
squares deconvolution \citep{Donati:1997fj} approach where we deconvolved
each spectrum against an ATLAS9 synthetic template \citep{Kurucz1992}
following the technique described by \citet{Zhou20}.

We collected five spectra of TOI-5427 between November
14, 2022 and February 11, 2023 using the NEID spectrograph
\citep{NEID,2016SPIE.9908E..6PH}, a stabilized fiber-fed optical
spectrograph on the WIYN 3.5-m telescope at Kitt Peak National Observatory
(KPNO).  We used the high-resolution (HR) mode that produces spectra with
$R \approx 110000$ with exposure times of 480 s.  The data were reduced
using v1.2.0 of the standard NEID Data Reduction Pipeline (NEID-DRP)
\footnote{\url{https://neid.ipac.caltech.edu/docs/NEID-DRP}}, which
derives radial velocities by cross-correlating the observed spectra
with a weighted stellar mask \citep{ELODIE,Pepe2002}.  The median radial
velocity precision attained was approximately 26 m/s.

\begin{deluxetable}{cccc}
\centering
\tabletypesize{\scriptsize}
\tablewidth{0pt}
\tablecaption{Ground-based Radial Velocities \label{tab:rvs}}
\tablehead{\colhead{BJD} &
\colhead{RV (km/s)} &
\colhead{$\sigma_{\text{RV}}$ (km/s)} &
\colhead{Source}}
\startdata
\multicolumn{4}{c}{TOI-5427 (UCAC4 515-012898)}\\
\hline
2459847.882200 & -44.961 & 0.023 & CHIRON\\
2459848.840030 & -48.368 & 0.110 & CHIRON\\
2459849.816380 & -38.673 & 0.033 & CHIRON\\
2459850.875790 & -28.476 & 0.095 & CHIRON\\
2459851.851630 & -30.545 & 0.052 & CHIRON\\
2459852.934530 & -42.431 & 0.038 & CHIRON\\
2459854.774330 & -42.442 & 0.069 & CHIRON\\
2459855.860460 & -30.129 & 0.059 & CHIRON\\
2459856.872970 & -28.896 & 0.036 & CHIRON\\
2459858.810920 & -47.816 & 0.051 & CHIRON\\
2459859.871500 & -44.323 & 0.065 & CHIRON\\
2459899.80424983 & -40.6134 & 0.0167 & NEID\\
2459898.84596773 & -30.1702 & 0.026 & NEID\\
2459897.87815973 & -29.9688 & 0.0304 & NEID\\
2459986.6052418 & -32.669 & 0.0262 & NEID\\
2459983.81974946 & -43.1964 & 0.0277 & NEID\\
\hline
\multicolumn{4}{c}{TOI-2533 (TYC 2010-124-1)}\\
\hline
2459278.939688 & 0.00000 &  0.09313 & TRES\\
2459295.873683 &  -12.98947 &  0.06914 & TRES\\
2459296.893338 &  -8.77543 &  0.09313 & TRES\\
2459298.965028 &  -0.02706 &  0.07976 & TRES\\
2459299.935610 &  1.81353 &  0.10406 & TRES\\
2459300.924987 &  -3.61235 &  0.14566 & TRES\\
2459301.855739 &  -12.74344 &  0.09697 & TRES\\
2459307.836527 &  -6.70705 &  0.10400 & TRES\\
2459309.818673 &  -10.75052 &  0.07288 & TRES\\
2459311.900542 &  -1.65152 &  0.07649 & TRES\\
2459312.931876 &  1.23157 &  0.08170 & TRES\\
\enddata
\end{deluxetable}

To fit Keplerian orbits to these ground-based radial velocity data
for TOI-2533 and TOI-5427, we use the \texttt{thejoker} Python package
\citep{2017ApJ...837...20P}.  We do not fit the TRES spectra for TOI-1712.
For each system, we use the default priors in \texttt{thejoker}: a
log-uniform prior in period set to be between the 1-$\sigma$ confidence
interval provided by TESS, a Gaussian prior in Doppler semiamplitude
described by $\sigma_K = 1$ km/s, and a Gaussian prior in systemic
velocity described by $\sigma_v = 1$ km/s.  We sample this prior 200000
times and conduct rejection sampling to obtain a set of initial parameters
for use in a \texttt{pymc3} \citep{2016ascl.soft10016S} MCMC simulation.
Using 10 chains, we generate a posterior distribution of 5000 samples
with a burn-in of 20000 samples.  The resulting posterior distributions
are both symmetric and unimodal.  As suggested by the \texttt{thejoker}
documentation\footnote{\url{https://thejoker.readthedocs.io/en/latest/}},
this unimodality suggests that these posterior samples are independent
and represent a unique solution without the need for convergence tests
due to the MCMC sampling method used by \texttt{thejoker}.  We then
take the 16th, 50th, and 84th percentiles of the resulting Doppler
semiamplitude and eccentricity posteriors to obtain the values and
uncertainties reported in Table \ref{tab:prop}.

\begin{figure*}
    \centering
    \includegraphics[width=\linewidth]{ 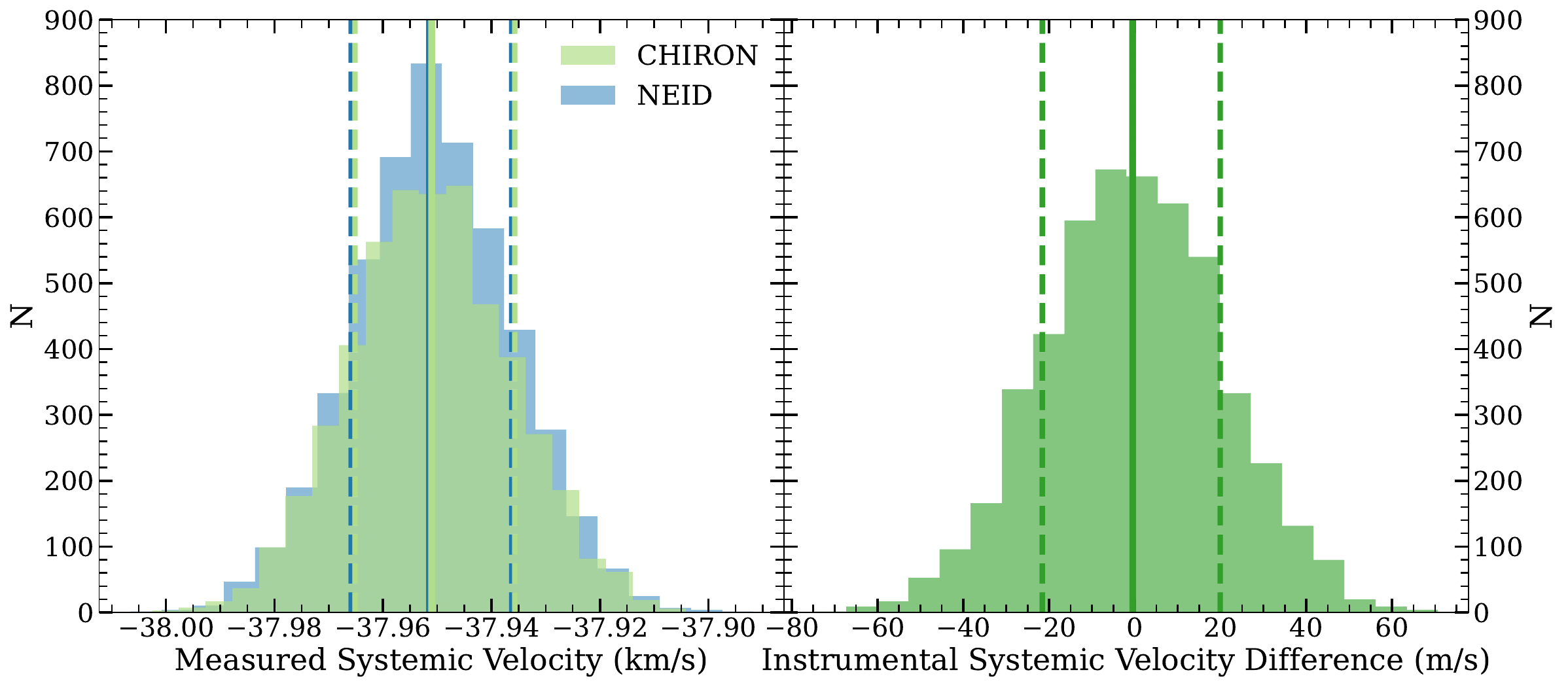}
    \caption{Comparison between CHIRON and NEID velocity zero-point
    offsets for TOI-5427.  We plot distribution medians as solid vertical
    lines and the 16th and 84th percentiles as dashed vertical lines.
    Left: Distributions of Monte Carlo samples from the \texttt{thejoker}
    Doppler solution systemic velocities after fitting to radial velocity
    measurements taken by CHIRON and NEID.  Right: Distribution of the
    differences between the two distributions in the left-hand panel.
    The resulting median offset is much smaller than the individual
    radial velocity uncertainties, suggesting that the data sets are
    consistent with each other.}
    \label{fig:offsets}
\end{figure*}

Zero-point offsets between different precision radial velocity
spectrographs with amplitudes of a few m/s are common.  We therefore
verify that any zero-point offsets have not affected our analysis
of TOI-5427.  We fit the radial velocities from each instrument with
\texttt{thejoker} using the same process described above.  We then
compare the systemic velocity distributions between the two in Figure
\ref{fig:offsets} and obtain 16th, 50th, and 84th percentiles of the
offset distribution by subtracting the systemic velocity posterior from
NEID from the systemic velocity posterior from CHIRON.  This results
in an offset of $-0.6^{+21.0}_{-20.5}$ m/s that is smaller than
the uncertainties of the individual radial velocity measurements.
This supports our decision to analyze both data sets without including
zero-point offsets between them.  We therefore combine the data from
CHIRON and NEID to simultaneously fit the ensemble of radial velocities
for TOI-5427.

\subsection{Determination of the Companions' Properties}\label{sec:montecarlo}

We solve the binary star mass function assuming that $\sin^3{i} \approx
1$ and re-parameterize Equation (\ref{binarymassfunction}) as a cubic
equation
\begin{equation} \label{eq:morecorrectradialvelocity}
   \left(\frac{M_{2}}{M_{1}}\right)^{3} - B \left(\frac{M_{2}}{M_{1}}\right)^{2} - 2 B\left(\frac{M_{2}}{M_{1}}\right) - B = 0,
\end{equation}
where the coefficient $B$ is defined as
\begin{equation}
   B = \frac{1}{2 \pi G} \frac{P K_{1}^{3} \left(1-e^{2} \right)^{3/2}}{M_{1}}.
\end{equation}
Taking the real root of Equation (\ref{eq:morecorrectradialvelocity})
results in a mass ratio that can then be used to determine companion
masses given primary masses.  Using the inclinations from each system's
30-minute cadence light curve \texttt{juliet} transit fit, we find that
the $\sin{i}$ corrections for the companions result in adjustments smaller
than 0.1\%.  The inclination uncertainties are therefore insignificant
in comparison to our mass estimate uncertainties.

We use the 30-minute cadence light curve \texttt{juliet}-inferred
radius ratio posteriors to calculate the secondary radii $R_{2}$ in
Jupiter radii.  In the TOI-1712 system, the flux from the eclipsing pair
is diluted by an unresolved secondary revealed by the TRES spectra.
We correct the dilution in the TOI-1712 light curve by incorporating
into a \texttt{juliet} transit fit the TESS $T$-band flux ratio from
our \texttt{isochrones} binary star fit posterior.  We first take each
point in the \texttt{isochrones} posterior and calculate the flux ratio
of the secondary to the primary
\begin{equation} \label{eq:flux}
   F = \frac{f_{\text{B}}}{f_{\text{Aa}}} = 10^{(T_{\text{Aa}} - T_{\text{B}})/2.5},
\end{equation}
where $T_{\text{Aa}}$ and $T_{\text{B}}$ are the TESS $T$-band magnitudes
of the primary and secondary.  We next calculate the dilution $D$ in
the light curve for each point
\begin{equation}\label{eq:dilution}
    D = \frac{1}{1 + F}.
\end{equation}
We then fit the 30-minute cadence light curve of TOI-1712 for each point
using the same procedure described in Section \ref{sec:lightcurves},
but with the dilution prior set to be the posterior point's calculated
dilution.  We obtain the median radius ratio from each point's individual
\texttt{juliet} posterior, and use these points as our full radius
ratio posterior for BD+45 1593.  We then use the same procedure as with
TYC 2010-124-1 and UCAC4 515-012898 to obtain the companion radius for
BD+45 1593.

We use a Monte Carlo simulation to infer the mass and radius uncertainties
for these three systems.  We generate normal distributions for each
\texttt{thejoker} Doppler analysis-derived (for TYC 2010-124-1 and UCAC4
515-012898) or Gaia/TESS-derived (for BD+45 1593) parameter.  The number
of Monte Carlo trials is equal to the lesser of the number of data points
in the mass and radius posterior distributions from \texttt{isochrones}
and the number of data points in the radius ratio posterior from
\texttt{juliet} (about 3000 samples).  For the posterior with more data
points, we randomly select a number of posterior points equal to the
number of data points in the smaller posterior without replacement.
For each set of points in the combined posteriors, we calculate a mass
and radius inference for the companion, resulting in distributions that
we characterize with their 16th, 50th, and 84th percentiles reported in
Table \ref{tab:prop}.  This approach captures the covariance between the
stellar parameters from the isochrone fit in order to generate an accurate
uncertainty for the companion mass and radius inferences.  A caveat
is that our reported uncertainties only include random uncertainties.
The inclusion of systematic uncertainties would likely result in larger
overall uncertainties \citep[e.g.,][]{Tayar2022}.

\begin{figure*}
    \centering
    \includegraphics[width=0.32\linewidth]{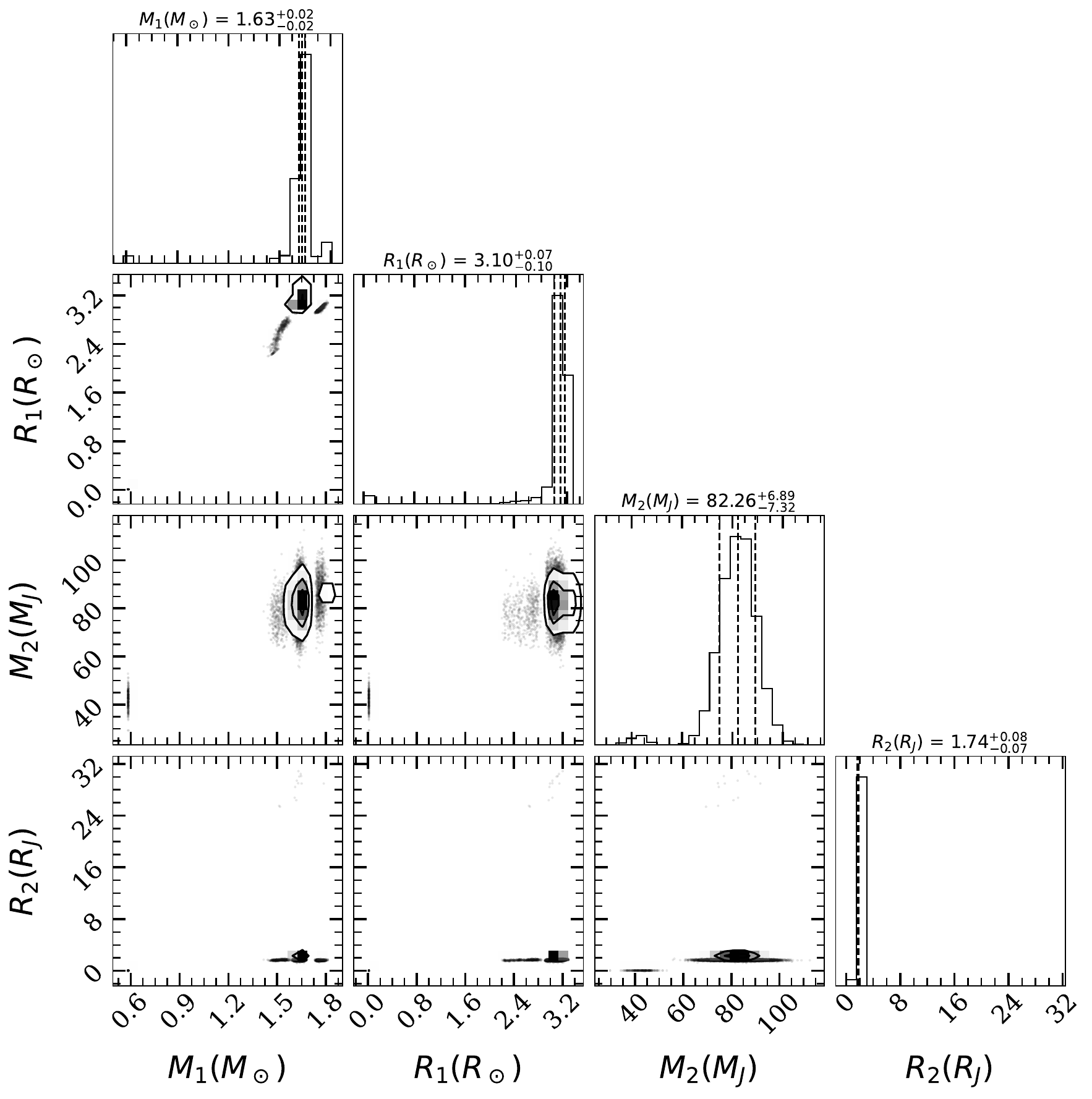}
    \includegraphics[width=0.32\linewidth]{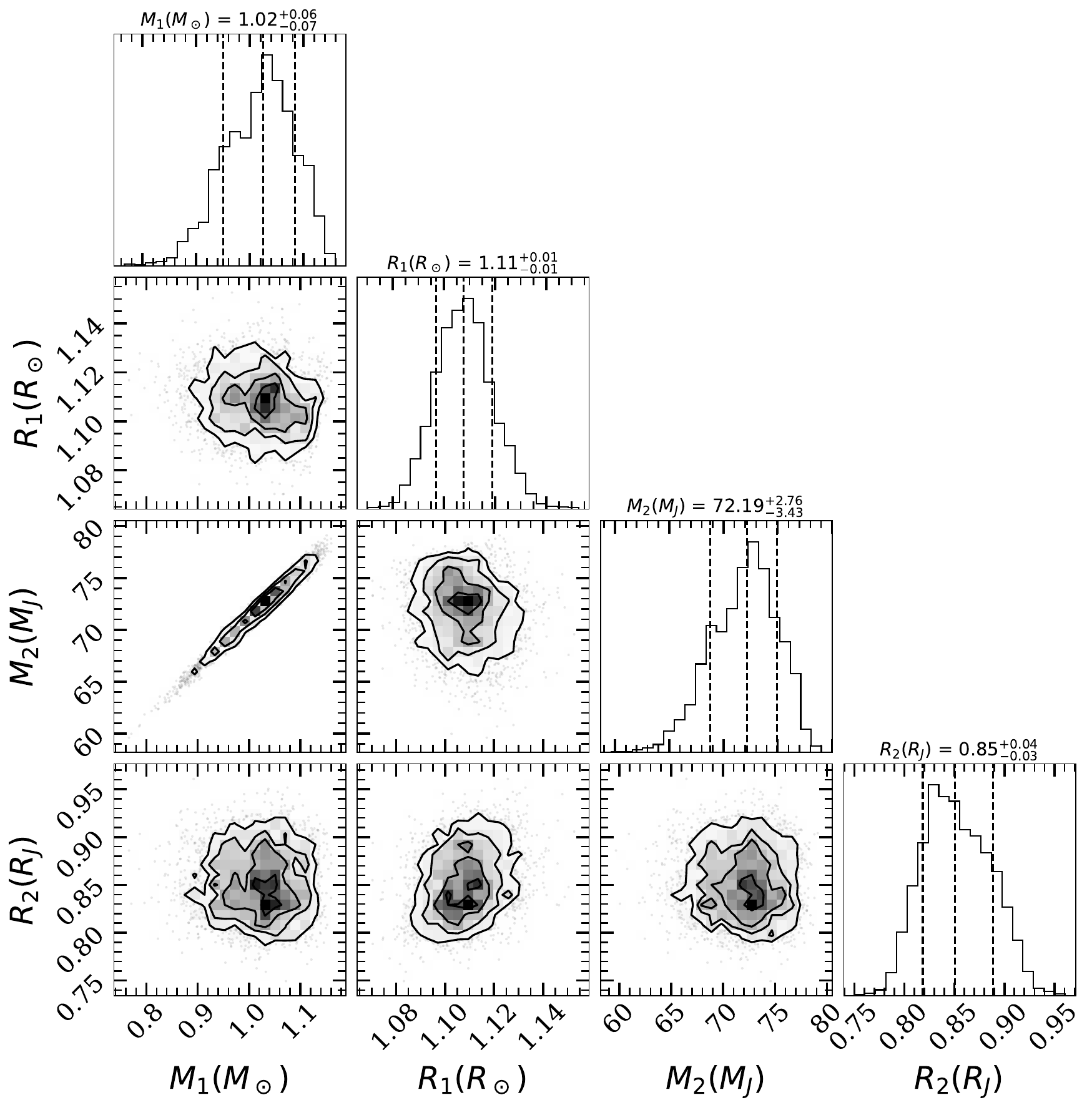}
    \includegraphics[width=0.32\linewidth]{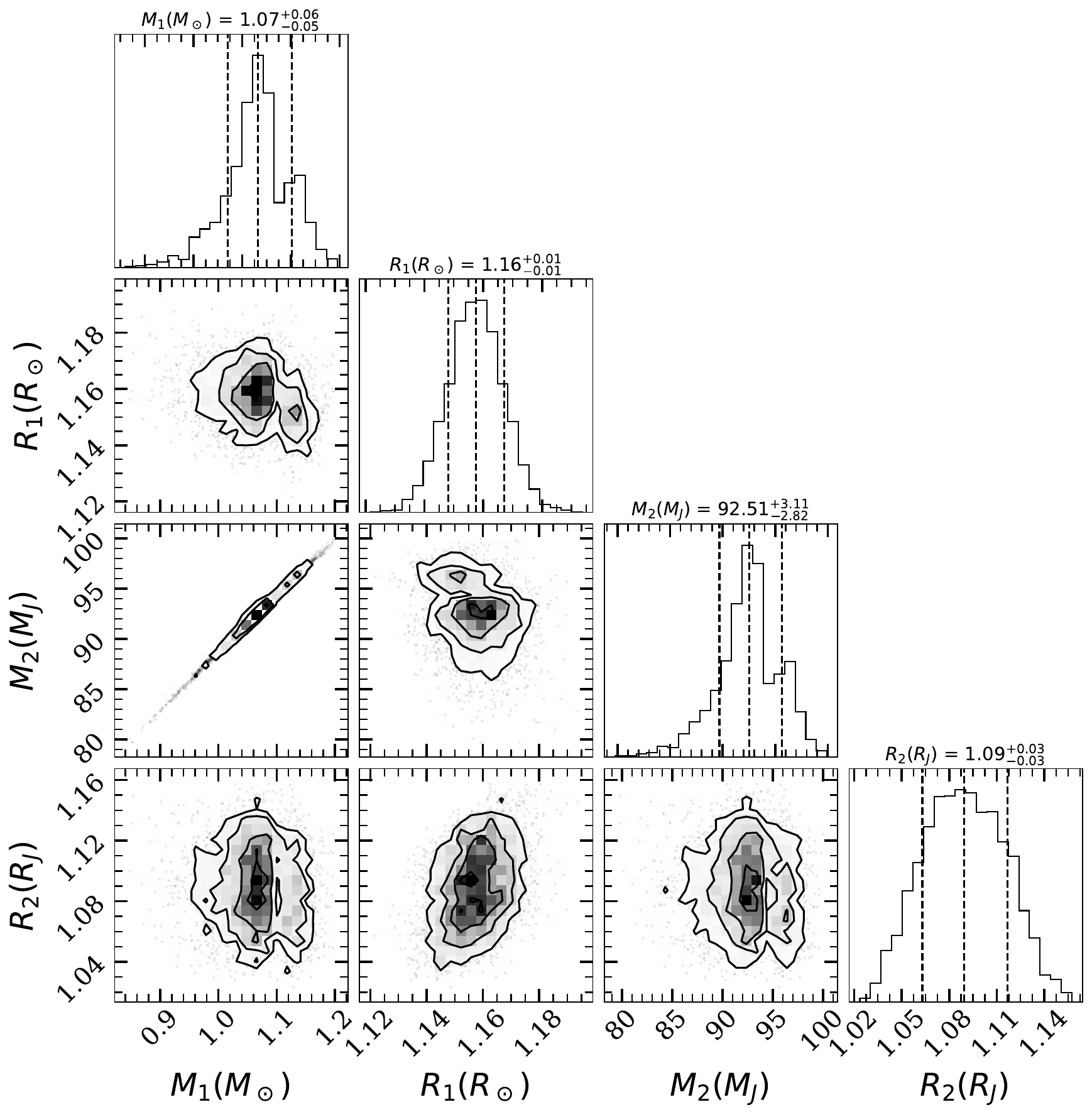}
    \caption{Corner plots showing unimodal posteriors for BD+45 1593
    (left), TYC 2010-124-1 (center), and UCAC4 515-012898 (right).}
    \label{fig:corner}
\end{figure*}

We account for contamination in the TESS light curve for UCAC4 515-012898
(TOI-5427), where the TESS contamination ratio is 26.3055\%.  This level
of light curve contamination corresponds to a radius underestimate
of 5.13\%, so we increase our inferred radius for UCAC4 515-012898 b
(TOI-5427) by 5.13\% to account for this effect.  BD+45 1593 (TOI-1712)
and TYC 2010-124-1 (TOI-2533) have TESS contamination ratios indicative
of radii underestimates that are smaller than our report uncertainties,
so we do not increase uncertainties in either case.  We provide corner
plots illustrating this process and visualizing the covariances for each
systems in Figure \ref{fig:corner}.

\begin{deluxetable*}{lcccc}
\centering
\tabletypesize{\scriptsize}
\tablewidth{0pt}
\tablecaption{Properties of the Objects and their Host Stars \label{tab:prop}}
\tablehead{\colhead{Parameter} &
\colhead{BD+45 1593} &
\colhead{TYC 2010-124-1} &
\colhead{UCAC4 515-012898} &
\colhead{Source}}
\startdata
\multicolumn{5}{c}{Identifiers}\\
\hline
TOI & 1712.01 & 2533.01 & 5427.01 & \citet{TOICatalog}\\
TIC & 67926921 &   418012030  &  52420398 &  \citet{TIC2018}\\
Gaia & 917823348536224768 &  1259922196651254272 & 3341401062125577088  & Gaia DR3 \\
2MASS & J08314698+4449564 &  J14103331+2625241  &  J05355241+1255340 & 2MASS\\
TYC & 2981-1719-1 &  2010-124-1  &  \nodata & Tycho-2\\
\hline
\multicolumn{5}{c}{Astrometry}\\
\hline
$\alpha$ & 127.945701 &  212.638795 & 83.968375 & Gaia DR3\\
$\delta$ & 44.832313 &   26.423432 & 12.925942 & Gaia DR3\\
$\alpha_{\text{J2000}}$ & 08h31m46.970745s &  14h10m33.327483s & 05h35m52.409305s & Gaia DR3\\
$\delta_{\text{J2000}}$ & 44d49m56.431574s &  26d25m24.046501s & 12d55m33.913407s & Gaia DR3\\
$\pi$ (mas) & $1.7653 \pm 0.0182$ &  $2.7078 \pm 0.0221$ & $2.7589 \pm 0.0158$ &  Gaia DR3\\
\hline
\multicolumn{5}{c}{Photometry}\\
\hline
FUV (mag) & $18.2406 \pm 0.0591$ &  \nodata  &  \nodata & GALEX\\
NUV (mag) & $13.9118 \pm 0.004$ &  \nodata  &  \nodata & GALEX\\
$B_{\text{T}}$ (mag) & $10.817 \pm 0.045$  &  $12.354 \pm 0.16$ & \nodata & Tycho-2 \\%
$V_{\text{T}}$ (mag) &   $10.485 \pm 0.04$7  &  $11.745 \pm 0.131$ &  \nodata & Tycho-2\\%
$G$ (mag) &  $10.2738 \pm 0.0017$ &  $12.0092\pm 0.0018$  & $12.0787\pm 0.0017$ & Gaia DR2\\
$T$ (mag) & $9.9841 \pm 0.0062$ & $11.6417 \pm 0.0069$ & $11.659 \pm 0.0061$ & TESS\\
$J$ (mag) & $9.541 \pm 0.0182$ &  $11.142 \pm 0.0242$  & $11.062 \pm 0.0201$ & 2MASS\\
$H$ (mag) & $9.374 \pm 0.0156$  & $10.872 \pm 0.0308$  &  $10.753 \pm 0.0176$ & 2MASS\\
$K_s$ (mag) & $9.353 \pm 0.016$ &  $10.8 \pm 0.019$  & $10.68 \pm 0.0172$ & 2MASS\\
$W1$ (mag) & $9.314 \pm 0.023$  &  $10.78 \pm 0.024$  &  $10.669 \pm 0.023$ & AllWISE\\
$W2$ (mag) & $9.327 \pm 0.021$ &  $10.798 \pm 0.022$  &  $10.701 \pm 0.022$ & AllWISE\\
$W3$ (mag) & $9.287 \pm 0.037$ &  $10.735 \pm 0.07$  &  $10.707 \pm 0.096$ & AllWISE\\
$A_V$ (mag) & $0.029^{+0.001}_{-0.016}$ &  $0.019^{+0.001}_{-0.021}$  &  $0.055^{+0.039}_{-0.039}$ & See Footnote\\ 
\hline
\multicolumn{5}{c}{Spectroscopy}\\
\hline
\texttt{vbroad} (km s$^{-1}$) & $	30 \pm 6$ &  \nodata & \nodata & Gaia DR3\\
\hline
\multicolumn{5}{c}{Single-lined Spectroscopic Binary Solution}\\
\hline
$K_1$ (km s$^{-1}$) & $7.69 \pm 0.61 $ & $7.39 \pm 1.10 $  &  $7.86 \pm 0.98$ & Gaia DR3\\
 & \nodata & 7.575 $\pm$ 0.046   &  9.841 $\pm$ 0.018 & Ground-Based\\
$e$ & $0.090 \pm 0.072$ &  $0.195 \pm 0.131$  &  $0.28 \pm 0.11$ & Gaia DR3\\
 & \nodata & $0.2473 \pm  0.0059$   &  $0.0332 \pm 0.0015$ & Ground-Based\\
$P$ (days) & $3.566276 \pm 0.0000044$ & $6.6847 \pm 0.0034$  &  $5.2374183 \pm 0.0000056$ & TESS \\
\hline
\multicolumn{5}{c}{Light Curve Transit Fit}\\
\hline
$i$ ($^\circ$) & $88.36^{+0.25}_{-0.25}$  & $88.01^{+1.34}_{-1.94}$  & $87.48^{+1.50}_{-1.65}$  & This work\\
$R_2$ / $R_\ast$ & $0.0564^{+0.0006}_{-0.0003}$ & $0.079^{+0.004}_{-0.003}$ & $0.097^{+0.002}_{-0.002}$ & This work\\
\hline
\multicolumn{5}{c}{Physical Properties}\\
\hline
T$_{\text{eff}}$ (K) & $6860^{+40}_{-30}$ & $6180^{+60}_{-60}$ & $6160^{+70}_{-40}$ & This Work \\
log \textit{g} & $3.67^{+0.03}_{-0.02}$ & $4.36^{+0.03}_{-0.03}$ & $4.34^{+0.03}_{-0.02}$ & This Work \\
$\left[ \mathrm{Fe/H} \right]$ & $-0.2^{+0.1}_{-0.1}$ & $-0.3^{+0.2}_{-0.2}$ & $-0.2^{+0.1}_{-0.1}$ & This Work \\
$M_\star$ ($M_\odot$) & $1.63^{+0.01}_{-0.02}$ &  $1.02^{+0.06}_{-0.07}$  &  $1.07^{+0.06}_{-0.05}$ & This Work \\
$R_\star$ ($R_\odot$) & $3.10^{+0.07}_{-0.10}$ & $1.11^{+0.01}_{-0.01}$ & $1.16^{+0.01}_{-0.01}$ & This Work \\
Age (Gyr) & $1.62^{+0.05}_{-0.03}$ &   $4.26^{+2.18}_{-1.74}$  &  $3.96^{+1.24}_{-1.53}$ & This Work \\
$M_2$ ($M_{\text{Jup}}$) &  $82^{+7}_{-7}$ ($^*$)  &  $72^{+3}_{-3}$ &  $93^{+3}_{-3}$ & This Work \\
$R_2$ ($R_{\text{Jup}}$) & $1.74^{+0.08}_{-0.07}$ &   $0.85^{+0.04}_{-0.03}$  &  $1.15^{+0.03}_{-0.03}$ & This Work \\
\enddata
\tablecomments{We derive the secondary mass for TYC 2010-124-1 and
UCAC4 515-012898 using ground-based radial velocity data from TRES
and both CHIRON and NEID, respectively.  We use \citet{Reddening_1}
and \citet{Reddening_2} for the $A_V$ extinctions.}
\end{deluxetable*}

\section{Discussion}\label{sec:disc}

Our synthesis of Gaia DR3 SB1 solutions and TESS transit
discoveries described in this article has revealed a new transiting
brown dwarf and two very low-mass eclipsing binary companions.
We confirm that TOI-2533.01 is a $72^{+3}_{-3}~M_{\text{Jup}} =
0.069^{+0.003}_{-0.003}~M_\odot$ transiting brown dwarf, placing
it in a class of only a few dozen such objects.  We include
in Table \ref{tab:list} an up-to-date list of known transiting
brown dwarfs.  We further confirm that TOI-5427.01 with a mass of
$93^{+3}_{-3}~M_{\text{Jup}} = 0.088^{+0.002}_{-0.002}~M_\odot$ is
one of the lowest-mass stars that models suggest is just above the
hydrogen-burning limit for its composition.  We validate TOI-1712.01 as an
inflated $82^{+7}_{-7}~M_{\text{Jup}} = 0.079^{+0.007}_{-0.007}~M_\odot$
very low-mass star in a hierarchical triple system.

The Gaia DR3 SB1 solutions for brown dwarfs and very low-mass stars are
consistent with the ground-based precision radial velocity-based SB1
solutions.  While the Gaia Doppler semiamplitude for TYC 2010-124-1 is
in good agreement with ground-based follow-up observations, it appears
that Gaia has underestimated by about 2 km/s the Doppler semiamplitude
of UCAC4 515-012898.  The Gaia DR3 SB1 solutions also suggest a
higher eccentricity for UCAC4 515-012898 and a slight phase shift
for TYC 2010-124-1.  We are unable to further investigate this issue,
as the individual Gaia radial velocity measurements are not publicly
available.  For these two systems, the agreement between the Gaia DR3
and ground-based precision radial velocity SB1 solutions suggests that
Gaia DR3 SB1 solutions are sufficiently precise to characterize massive
brown dwarfs and low-mass stars.

It seems possible that the quality of the Gaia DR3 SB1 solution is
proportional to the number of ``good'' radial velocity observations
available by the time of Gaia DR3.  The brown dwarf TYC 2010-124-1 b has
47 ``good'' Gaia DR3 radial velocities, while the very low-mass stars
BD+45 1593 Ab and UCAC4 515-012898 b only have 18 and 25, respectively.
We suggest that the number of ``good'' Gaia radial velocities may be
a useful metric for assessing future Gaia SB1 solutions.  The fact that
BD+45 1593 is an SB2 system may have affected Gaia's SB1 fit to its radial
velocity variations.  BD+45 1593 A is both hot and rotating quickly, and
all of these facts make it a challenging target for Doppler observations.

\begin{figure*}
    \centering
    \includegraphics[width=\linewidth]{ 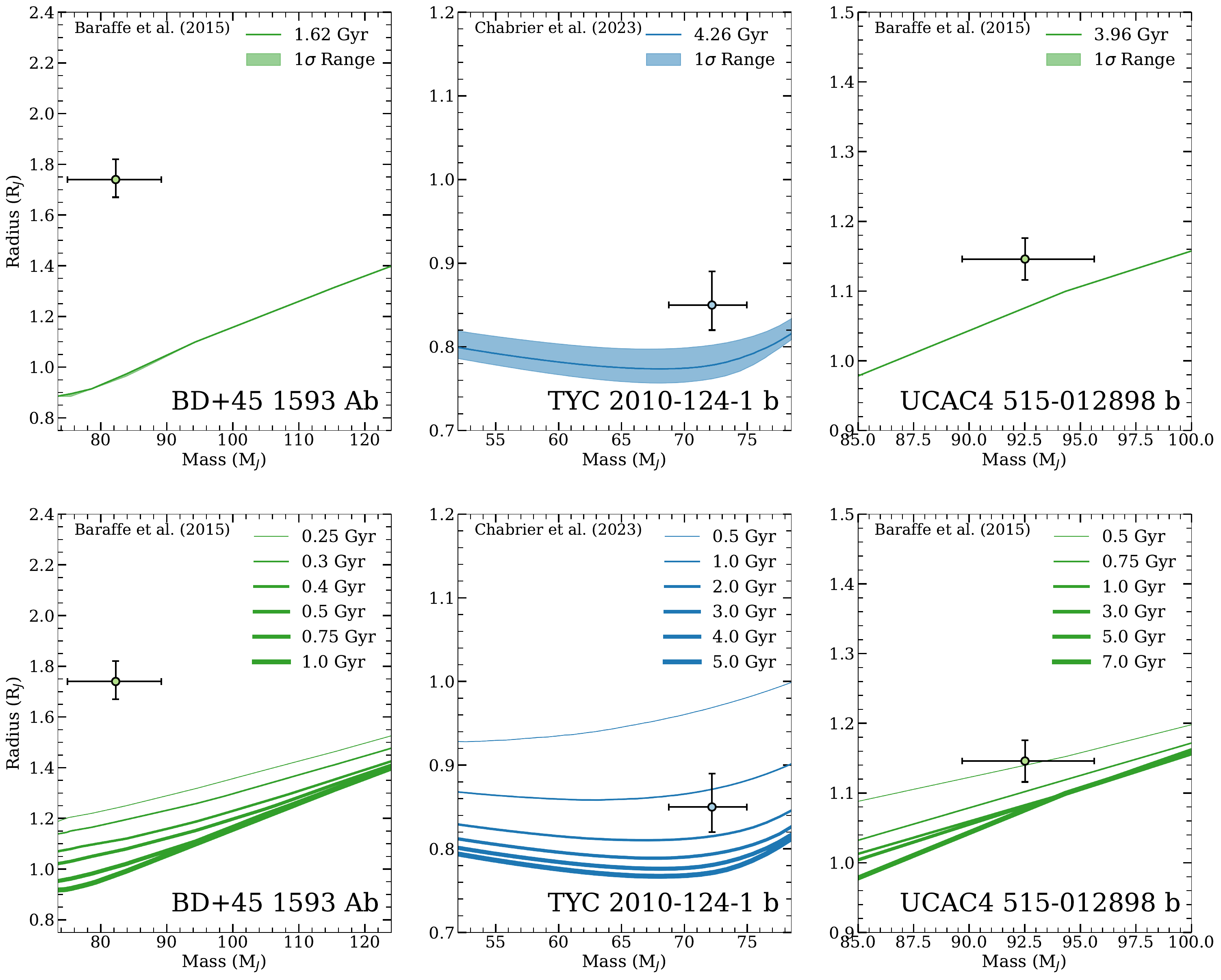}
    \caption{Comparisons between predicted and observed masses and radii.
    Top: mass--radius relations assuming our isochrone-inferred system
    ages. We plot the mass--radius relation at each system's age as
    the solid line and indicate its 1-$\sigma$ uncertainty as the
    shaded region.  Bottom: mass--radius relations assuming a range
    of ages. Thicker lines refer to older ages.  BD+45 1593 Ab has a
    significantly larger radius than predicted by the \citet{BHAC15}
    theoretical models.  UCAC4 515-012898 b is slightly bigger than
    predicted by the \citet{BHAC15} theoretical models as is frequently
    observed for low-mass stars. TYC 2010-124-1 b is consistent with the
    \citet{2023A&A...671A.119C} models. The models we use assume solar
    metallicity, while our systems have slightly subsolar metallicities.
    However, these differences in metallicity are too small to produce 
    a radius offset as large as we observe for BD+45 1593 Ab.  Indeed,
    at the low-mass end of the MIST isochroe grid at $M_{\ast} \approx
    0.11~M_{\odot} \approx 110~M_{\text{Jup}}$ a 0.2 dex decrease
    in $[\text{Fe/H}]$ corresponds to no more than a 5\% decrease in
    stellar radius.}
    \label{fig:isochrones}
\end{figure*}

We plot our inferred masses and radii for the three secondaries in Figure
\ref{fig:isochrones}.  While we find that our mass and radius inferences
for TYC 2010-124-1 b are consistent with the \citet{2023A&A...671A.119C}
theoretical models for brown dwarfs, the radius of UCAC4 515-012898 b
is about 6\% larger than predicted by the \citet{BHAC15} theoretical
models for isolated stars of the same mass, as is often the case for
low-mass stars.  In contrast, the radius of BD+45 1593 Ab $R_{2} =
1.74^{+0.08}_{-0.07}~R_{\text{Jup}}$, is more than 80\% larger than the
radius $R \approx 1.0~R_{\text{Jup}}$ predicted by the \citet{BHAC15}
models for an isolated star with similar mass, composition, and age.
Because the MIST and PAdova TRieste Stellar Evolutionary Code
\citep[PARSEC -][]{PARSEC,PARSEC2} isochrone grids do not extend to
these objects' mass ranges, we are unable to use them for comparison.
Given the precisions of our mass, age, composition, and radius inferences,
we find at over 9-$\sigma$ significance that BD+45 1593 Ab is inflated.
We note that this estimate accounts for dilution from BD+45 1593 B,
and in any case dilution typically would lead to an underestimated radius.

Observations of radii about 10\% larger than predicted by
theoretical models appear to be ubiquitous among K and M dwarfs
\citep{Kraus2011,Birkby2012}.  This larger observed radii correspond
to observed temperatures typically 5\% cooler than predicted by models,
though this effect is poorly studied due to the difficulty of determining
late-type stellar temperatures \citep{Rib08,Morales2009a,Tor13}.  Magnetic
fields are the most-favored explanation for this ``radius inflation
problem'' (and its parallel ``temperature suppression problem''), as they
can increase the radii of stars yet are rarely included in stellar models.
Strong magnetic fields are responsible for the significant starspot
coverage of late-type dwarfs and may inhibit convective energy transport
\citep{Feiden2013}.  The net result of these two physical mechanisms is
an increase in the photospheric radii of K and M dwarfs.

To empirically correct for radius underestimates, attempts have
been made in the past to relate magnetic activity with proxies like
rotation \citep{Somers2017,2019ApJ...879...39J}, X-ray emission
\citep{LopezMorales2007}, and H$\alpha$ emission \citep{Stassun2012}.
On the other hand, \citet{Mann2015b} found no correlation between radius
discrepancy and either H$\alpha$ or X-ray emission.  While models have had
some success including magnetic fields in the context of X-ray emission
\citep{2014ApJ...787...70M}, the magnetic field strengths required in this
context might not be realistic \citep{Feiden2014a}.  More quantitative
predictions from \citet{2016ApJ...818..189B} provided an upper bound for
the possible strength of these magnetic fields via magnetohydrodynamic
simulations, and \citet{2015MNRAS.448.2019M} provided a constraint based
on lithium abundance.  While this remains an active area of research, it
appears that magnetic fields contribute to radius inflation in low-mass
stars without accounting for all aspects of the radius inflation problem.

The eventual solution of the radius inference problem is important,
as accurate and precise radii for low-mass stars are required for the
inference of accurate and precise radii for the apparently common planets
transiting low-mass stars \citep{Cassan2012}.  As low-mass stars are also
the most common type of star in the Galaxy \citep{2010AJ....139.2679B},
we expect that a sizable portion of the Milky Way's exoplanet population
orbits low-mass stars.  Because of their small separations and high
transit probabilities, habitable zone terrestrial planets are easier to
find orbiting low-mass stars than solar-type stars.  Low-mass stars are
therefore the preferred target for habitable transiting planet search
programs \citep{2003ApJ...594..533G} like MEarth \citep{Nutzman2008},
the TRAnsiting Planets and PlanetesImals Small Telescope \citep[TRAPPIST
-][]{2011Msngr.145....2J}, the Next-Generation Transit Survey \citep[NGTS
-][]{2016Msngr.165...10W}, and TESS \citep{2014SPIE.9143E..20R}.
Unaccounted for, the radius inflation problem would lead to underestimated
radii for planets transiting M dwarfs.  An underestimated radius could
cause a non-terrestrial, uninhabitable planet to be mistakenly classified
as a terrestrial, possibly habitable planet.

We propose that the intense instellation experienced by BD+45 1593 Ab
is primarily responsible for its inflated radius.  It has been suggested
that in binary systems with two very low-mass stars, instellation-driven
radius inflation is limited to 5\% \citep{2017A&A...601A..75L}.  In this
situation though, BD+45 1593 Ab is experiencing 500 times more flux than
in situation investigated by \citet{2017A&A...601A..75L}.  The increased 
surface temperature resulting from this instellation could decrease
the temperature gradient of BD+45 1593 Ab near its surface, thereby
limiting the ability of convection to transport energy in the same
manner as magnetic fields.  This diminished efficiency of convective
energy transport would then lead to the inflated radius of BD+45 
1593 Ab.  Indeed, the photospheric temperature of an isolated star with
similar properties to BD+45 1593 Ab is expected to be $T_{\text{eff}}
\approx 2300$  \citep{BHAC15}.  Give its instellation, the ``day 
side'' equilibrium temperature of  BD+45 1593 Ab could be has high as 
$T_{\text{eq}} \approx 4400$ K assuming a Bond albedo $A_{\text{B}} = 
0$ and no redistribution of heat.

Tidal effects may also play a role in the inflated radius of BD+45 1593
Ab, perhaps in a process analogous to the shear mixing that may occur
in massive binaries \citep{2013A&A...556A.100S}.  The fully convective
nature of BD+45 1593 Ab would make shear mixing inefficient, and its small
eccentricity likely minimizes the importance of tidal dissipation for its
inflated radius.  The probable fast rotation of BD+45 1593 Ab resulting
from its circular orbit, short circularization time, and therefore likely 
spin--orbit synchronization could contribute to its inflated radius,
though the relationship between rotation and radius inflation is unclear
\citep{Jackson2019}.  Consequently, we conclude that instellation is
the best candidate to explain the inflated radius of BD+45 1593 Ab.

We searched the light curves of TOI-1712 and TOI-5427 for secondary 
eclipses perfectly opposite the transits of TOI-1712.01 and TOI-5427.01 
as expected for circular orbits.  We were unable to identify a secondary
eclipse in either case.  This is not surprising, as the detection of a
secondary eclipse would have eliminated these objects from consideration
for inclusion in the TOI catalog.  Given the precision of the TESS data
for TOI-1712 and TOI-5427 and the expected secondary eclipse durations 
of TOI-1712.01 and TOI-5427.01, we would have been able to identify
secondary eclipses above 0.003\% for TOI-1712 and above 0.01\% for TOI-5427.
This lack of a secondary eclipse detection does not constrain the ``day
side'' temperature for UCAC4 515-012898 b, as we expect its photospheric
temperature to have been undetectable given our limits.  On the other
hand, the non-detection of a secondary eclipse for  TOI-1712.01 limits
the ``day side'' temperature of BD+45 1593 Ab to less than 3200 K.
The ``day side'' of BD+45 1593 Ab could be as hot as 4400 K if its
albedo is zero and heat is not redistributed, so our non-detection of
a secondary eclipse implies that either its albedo is nonzero or the 
redistribution of heat is efficient.


We compare TYC 2010-124-1 b to the \citet{2023A&A...671A.119C}
brown dwarf isochrones in Figure \ref{fig:isochrones}.
Isolated brown dwarfs cool and contract to settle into
a typical mass range of between 0.7 and 1.4 Jupiter radii
\citep{corot3b,2020AJ....159..151S,2020AJ....160...53C} after at least
100 Myr.  Objects younger than this tend to be larger \citep{David2019,
2006Natur.440..311S}.   According to isolated brown dwarf evolution models
\citep[e.g.,][]{2003A&A...402..701B,2008ApJ...689.1327S,2020A&A...637A..38P,2021ApJ...920...85M,2023MNRAS.519.5177C},
the rate of this contraction should decrease over time.  However,
the details of this process are unclear \citep{2023MNRAS.519.5177C}.
We find that the radius of TYC 2010-124-1 b is 9\% bigger than the
model prediction for an isolated brown dwarf, but this offset is only
significant at the 1.9-$\sigma$ level.

The radius inflation we observe in BD+45 1593 Ab potentially parallels
the processes responsible for the inflation of some hot Jupiters
\citep[e.g.,][]{Gru16,Gru17,Gru22,Wit22}.  Hot Jupiters orbiting evolved
stars are sometimes observed with larger radii than hot Jupiters orbiting
main sequence stars \citep{Gru19}.  In this case, the inflated radii of
mature hot Jupiters orbiting evolved stars have been ``reinflated'' due
to rapid transport of energy deposited into their surface layers by their
evolved host star into their interiors \citep[e.g.,][]{Lop16,Tho18,Tho21}.
Intense instellation may therefore be capable of inflating radii from
the giant planet regime up to the very low-mass star regime.

Because theoretical models have struggled to explain the data that is
currently available, a much larger sample of massive brown dwarfs and
very low-mass stars with precisely inferred radii relative to their
solar-type primaries could lead to useful empirical radius calibrations.
Our approach that combines Gaia SB1 solutions with transit discoveries
from Kepler, K2, TESS, and the future PLAnetary Transits and Oscillations
(PLATO) mission has the potential to precisely characterize over 400
brown dwarfs and low-mass stars.  Such a catalog of radii precisely
inferred relative to the radii of their primaries could be the basis of
future empirical brown dwarf and low-mass star mass--radius relations.
The masses inferred in this way using the Doppler technique with
a known inclination are more precise and likely more accurate than
other currently employed methods of mass inference such as astrometry
or atmospheric fitting\footnote{Different mass inference methods have
yielded inconsistent mass results for the first confirmed brown dwarf
Gliese 229 B \citep{1996Sci...272.1919M,2020AJ....160..196B}}. As transit
surveys typically focus on low-mass stars to find small-radius planets,
these empirical stellar radii could lead to better planet radii which
would lead to better atmospheric and interior structure inferences for
terrestrial planets like those in the TRAPPIST-1 system.

\begin{longrotatetable}
\begin{deluxetable}{ccccccccccc}
\tabletypesize{\scriptsize}
\tablewidth{0pt}
\tablehead{\colhead{Name} &
\colhead{Simbad Name} &
\colhead{$P$} &
\colhead{$M_{\text{BD}}$} &
\colhead{$R_{\text{BD}}$} &
\colhead{Eccentricity} &
\colhead{$M_{\ast}$} &
\colhead{$R_{\ast}$} &
\colhead{$T_{\text{eff}}$} &
\colhead{[Fe/H]} &
\colhead{Reference} \\
\colhead{} &
\colhead{} &
\colhead{(days)} &
\colhead{($M_{\text{Jup}}$)} &
\colhead{($R_{\text{Jup}}$)} &
\colhead{} &
\colhead{($M_{\odot}$)} &
\colhead{($R_{\odot}$)} &
\colhead{(K)} &
\colhead{} &
\colhead{}}
\tablecaption{Literature Sample of Transiting Brown Dwarfs as of June
2023 \label{tab:list}}
\startdata
\hline
TYC 2010-124-1 b & TYC 2010-124-1 & $6.6847^{+0.0034}_{-0.0034}$ & $72^{+3}_{-3}$ & $0.85^{+0.04}_{-0.03}$ & $0.2473^{+0.0059}_{-0.0059}$  & $1.02^{+0.06}_{-0.07}$ & $1.11^{+0.01}_{-0.01}$ & $6180^{+60}_{-60}$ & $-0.3^{+0.2}_{-0.2}$ &  This Work \\
TOI-2119 b & StM 274 & 7.201 & $64.4^{+2.3}_{-2.2}$ & $1.08^{+0.03}_{-0.03}$ & $0.337^{+0.0019}_{-0.00064}$ & $0.525^{+0.02}_{-0.021}$ & $0.5^{+0.015}_{-0.015}$ & $3621^{+48}_{-46}$ & $0.055^{+0.084}_{-0.077}$ & 1 \\
TOI-2521 b & TYC 5362-881-1 & 5.563 & $77.5^{+3.3}_{-3.3}$ & $1.01^{+0.04}_{-0.04}$ & $< 0.035 $ & $1.1^{+0.07}_{-0.07}$ & $1.77^{+0.068}_{-0.068}$ & $5600^{+100}_{-100}$ & $-0.3^{+0.3}_{-0.3}$ & 2 \\
TOI-2336 b & TYC 7317-698-1 & 7.712 & $69.9^{+2.3}_{-2.3}$ & $1.05^{+0.04}_{-0.04}$ & $0.01^{+0.006}_{-0.005}$ & $1.41^{+0.08}_{-0.08}$ & $1.781^{+0.059}_{-0.059}$ & $6550^{+100}_{-100}$ & $0^{+0.3}_{-0.3}$ & 2 \\
TOI-2543 b & TYC 229-654-1 & 7.543 & $67.62^{+3.45}_{-3.45}$ & $0.95^{+0.09}_{-0.09}$ & $0.009^{+0.003}_{-0.002}$ & $1.29^{+0.08}_{-0.08}$ & $1.86^{+0.15}_{-0.15}$ & $6060^{+82}_{-82}$ & $-0.28^{+0.1}_{-0.1}$ & 3 \\
TOI-1982 b & BD-22 3669 & 17.172 & $65.85^{+2.75}_{-2.72}$ & $1.08^{+0.04}_{-0.04}$ & $0.272^{+0.014}_{-0.014}$ & $1.41^{+0.08}_{-0.08}$ & $1.51^{+0.05}_{-0.05}$ & $6325^{+110}_{-110}$ & $-0.1^{+0.09}_{-0.09}$ & 3 \\
TOI-629 b & HD 44717 & 8.718 & $66.98^{+2.96}_{-2.95}$ & $1.11^{+0.05}_{-0.05}$ & $0.298^{+0.008}_{-0.008}$ & $2.16^{+0.13}_{-0.13}$ & $2.37^{+0.11}_{-0.11}$ & $9100^{+200}_{-200}$ & $0.1^{+0.15}_{-0.15}$ & 3 \\
TOI-1994 b & HD 298656 & 4.03 & $22$ & $1.28$ & $0.29$ & $\cdots$ & $\cdots$ & $7550$ & $\cdots$ & 4 \\
GPX-1 b & UCAC4 731-024744 & 1.75 & $19.7^{+1.6}_{-1.6}$ & $1.47^{+0.10}_{-0.10}$ & $\cdots$ & $1.68^{+0.10}_{-0.10}$ & $1.56^{+0.10}_{-0.10}$ & $7000^{+200}_{-200}$ & $\cdots$ & 5 \\
TOI-746 b & TOI-746 & 10.98 & $82.2^{+4.9}_{-4.4}$ & $0.95^{+0.09}_{-0.06}$ & $0.199^{+0.003}_{-0.003}$ & $0.98^{+0.06}_{-0.06}$ & $0.957^{+0.051}_{-0.051}$ & $5593^{+215}_{-215}$ & $0.01^{+0.29}_{-0.29}$ & 6 \\
TOI-587 b & HD 74162 & 8.04 & $81.1^{+7.1}_{-7.0}$ & $1.32^{+0.07}_{-0.06}$ & $0.051^{+0.049}_{-0.036}$ & $2.32^{+0.14}_{-0.14}$ & $2.031^{+0.092}_{-0.092}$ & $10400^{+300}_{-300}$ & $0.07^{+0.12}_{-0.12}$ & 6 \\
TOI-148 b & TOI-148 & 4.87 & $77.1^{+5.8}_{-4.6}$ & $0.81^{+0.05}_{-0.06}$ & $0.005^{+0.006}_{-0.004}$ & $1.03^{+0.06}_{-0.06}$ & $1.192^{+0.068}_{-0.068}$ & $5836^{+286}_{-286}$ & $-0.28^{+0.28}_{-0.28}$ & 6  \\
NGTS-19 b & NGTS-19 & 17.8397 & $69.5^{+5.7}_{-5.4}$ & $1.034^{+0.055}_{-0.053}$ & $0.3767^{+0.0061}_{-0.0061}$ & $0.807^{+0.038}_{-0.043}$ & $0.896^{+0.04}_{-0.035}$ & $4716^{+39}_{-28}$ & $0.11^{+0.074}_{-0.07}$ & 7 \\
TOI-1278 B & TOI-1278 & 14.476 & $18.5^{+0.5}_{-0.5}$ & $1.09^{+0.24}_{-0.2}$ & $0.013^{+0.004}_{-0.004}$ & $0.55^{+0.02}_{-0.02}$ & $0.573^{+0.012}_{-0.012}$ & $3799^{+42}_{-42}$ & $-0.01^{+0.28}_{-0.28}$ & 8 \\
TOI-811 b & TOI-811 & 25.166 & $59^{+13}_{-8.6}$ & $1.26^{+0.06}_{-0.06}$ & $0.509^{+0.075}_{-0.075}$ & $1.32^{+0.07}_{-0.07}$ & $1.27^{+0.09}_{-0.09}$ & $6107^{+77}_{-77}$ & $0.4^{+0.09}_{-0.09}$ & 9 \\
TOI-852 b & TOI-852 & 4.946 & $53.7^{+1.4}_{-1.4}$ & $0.83^{+0.04}_{-0.04}$ & $0.004^{+0.004}_{-0.004}$ & $1.32^{+0.05}_{-0.05}$ & $1.72^{+0.04}_{-0.04}$ & $5768^{+84}_{-84}$ & $0.33^{+0.09}_{-0.09}$ & 9 \\
HATS-70 b & HATS-70 & 1.888 & $12.9^{+1.8}_{-1.8}$ & $1.38^{+0.08}_{-0.08}$ & $< 0.18$ & $1.78^{+0.12}_{-0.12}$ & $1.88^{+0.07}_{-0.07}$ & $7930^{+820}_{-820}$ & $0.04^{+0.11}_{-0.11}$ & 10 \\
KELT-1 b & KELT-1 & 1.218 & $27.4^{+0.9}_{-0.9}$ & $1.12^{+0.04}_{-0.04}$ & $0.01^{+0.01}_{-0.01}$ & $1.34^{+0.06}_{-0.06}$ & $1.47^{+0.05}_{-0.05}$ & $6516^{+49}_{-49}$ & $0.05^{+0.08}_{-0.08}$ & 11 \\
NLTT 41135 b & LP 563-38 & 2.889 & $33.7^{+2.8}_{-2.8}$ & $1.13^{+0.27}_{-0.27}$ & $< 0.02$ & $0.19^{+0.03}_{-0.03}$ & $0.21^{+0.02}_{-0.02}$ & $3230^{+130}_{-130}$ & $-0.25^{+0.25}_{-0.25}$ & 12 \\
LHS 6343 c & G 205-57 & 12.713 & $62.9^{+2.3}_{-2.3}$ & $0.83^{+0.02}_{-0.02}$ & $0.056^{+0.032}_{-0.032}$ & $0.37^{+0.01}_{-0.01}$ & $0.38^{+0.01}_{-0.01}$ & $\cdots$ & $0.02^{+0.19}_{-0.19}$ & 13 \\
LP 261-75 b & LP 261-75 & 1.882 & $68.1^{+2.1}_{-2.1}$ & $0.9^{+0.02}_{-0.02}$ & $<0.007$ & $0.3^{+0.02}_{-0.02}$ & $0.31^{+0.01}_{-0.01}$ & $3100^{+50}_{-50}$ & $\cdots$ & 14 \\
WASP-30 b & WASP-30 & 4.157 & $62.5^{+1.2}_{-1.2}$ & $0.95^{+0.03}_{-0.03}$ & 0 (adopted) & $1.25^{+0.04}_{-0.04}$ & $1.4^{+0.03}_{-0.03}$ & $6202^{+51}_{-51}$ & $0.08^{+0.1}_{-0.1}$ & 15 \\
WASP-128 b & WASP-128  & 2.209 & $37.2^{+0.9}_{-0.9}$ & $0.94^{+0.02}_{-0.02}$ & $< 0.007$ & $1.16^{+0.04}_{-0.04}$ & $1.15^{+0.02}_{-0.02}$ & $5950^{+50}_{-50}$ & $0.01^{+0.12}_{-0.12}$ & 16 \\
CoRoT-3 b & CoRoT-3  & 4.257 & $21.7^{+1}_{-1}$ & $1.01^{+0.07}_{-0.07}$ & 0 (adopted) & $1.37^{+0.09}_{-0.09}$ & $1.56^{+0.09}_{-0.09}$ & $6740^{+140}_{-140}$ & $-0.02^{+0.06}_{-0.06}$ & 17 \\
CoRoT-15 b & CoRoT-15  & 3.06 & $63.3^{+4.1}_{-4.1}$ & $1.12^{+0.3}_{-0.3}$ & 0 (adopted) & $1.32^{+0.12}_{-0.12}$ & $1.46^{+0.31}_{-0.31}$ & $6350^{+200}_{-200}$ & $0.1^{+0.2}_{-0.2}$ & 18 \\
CoRoT-33 b & CoRoT-33  & 5.819 & $59^{+1.8}_{-1.8}$ & $1.1^{+0.53}_{-0.53}$ & $0.07^{+0.002}_{-0.002}$ & $0.86^{+0.04}_{-0.04}$ & $0.94^{+0.14}_{-0.14}$ & $5225^{+80}_{-80}$ & $0.44^{+0.1}_{-0.1}$ & 19 \\
Kepler-39 b & Kepler-39  & 21.087 & $20.1^{+1.3}_{-1.3}$ & $1.24^{+0.1}_{-0.1}$ & $0.112^{+0.057}_{-0.057}$ & $1.29^{+0.07}_{-0.07}$ & $1.4^{+0.1}_{-0.1}$ & $6350^{+100}_{-100}$ & $0.1^{+0.14}_{-0.14}$ & 20 \\
KOI-189 b & Kepler-486  & 30.36 & $78^{+3.4}_{-3.4}$ & $1^{+0.02}_{-0.02}$ & $0.275^{+0.004}_{-0.004}$ & $0.76^{+0.05}_{-0.05}$ & $0.73^{+0.02}_{-0.02}$ & $4952^{+40}_{-40}$ & $-0.07^{+0.12}_{-0.12}$ & 21 \\
KOI-205 b & Kepler-492  & 11.72 & $39.9^{+1}_{-1}$ & $0.81^{+0.02}_{-0.02}$ & $< 0.031 $ & $0.92^{+0.03}_{-0.03}$ & $0.84^{+0.02}_{-0.02}$ & $5237^{+60}_{-60}$ & $0.14^{+0.12}_{-0.12}$ & 22 \\
KOI-415 b & KOI-415  & 166.788 & $62.1^{+2.7}_{-2.7}$ & $0.79^{+0.12}_{-0.12}$ & $0.689^{+0.001}_{-0.001}$ & $0.94^{+0.06}_{-0.06}$ & $1.15^{+0.15}_{-0.15}$ & $5810^{+80}_{-80}$ & $-0.24^{+0.11}_{-0.11}$ & 23 \\
EPIC 201702477 b & UCAC2 33047398  & 40.737 & $66.9^{+1.7}_{-1.7}$ & $0.76^{+0.07}_{-0.07}$ & $0.228^{+0.003}_{-0.003}$ & $0.87^{+0.03}_{-0.03}$ & $0.9^{+0.06}_{-0.06}$ & $5517^{+70}_{-70}$ & $-0.16^{+0.05}_{-0.05}$ & 24 \\
EPIC 212036875 b & TYC 1400-1873-1  & 5.17 & $52.3^{+1.9}_{-1.9}$ & $0.87^{+0.02}_{-0.02}$ & $0.132^{+0.004}_{-0.004}$ & $1.29^{+0.07}_{-0.07}$ & $1.5^{+0.03}_{-0.03}$ & $6238^{+60}_{-60}$ & $0.01^{+0.1}_{-0.1}$ & 25 \\
AD 3116 b & Cl* NGC 2632 HSHJ 430  & 1.983 & $54.2^{+4.3}_{-4.3}$ & $1.02^{+0.28}_{-0.28}$ & $0.146^{+0.024}_{-0.024}$ & $0.28^{+0.02}_{-0.02}$ & $0.29^{+0.08}_{-0.08}$ & $3200^{+200}_{-200}$ & $0.16^{+0.1}_{-0.1}$ & 26 \\
CWW 89 Ab & UCAC4 366-166973  & 5.293 & $36.84^{+0.97}_{-0.97}$ & $0.94^{+0.03}_{-0.03}$ & $0.1929^{+0.002}_{-0.002}$ & $1.01^{+0.04}_{-0.04}$ & $1.01^{+0.03}_{-0.03}$ & $5850^{+85}_{-85}$ & $0.03^{+0.08}_{-0.08}$ & 27 \\
RIK 72 b & UGCS J160339.22-185129.4  & 97.76 & $59.2^{+6.8}_{-6.8}$ & $3.1^{+0.31}_{-0.31}$ & $0.146^{+0.012}_{-0.012}$ & $0.44^{+0.04}_{-0.04}$ & $0.96^{+0.1}_{-0.1}$ & $3349^{+142}_{-142}$ & $\cdots$ & 28 \\
TOI-503 b & BD+13 1880  & 3.677 & $53.7^{+1.2}_{-1.2}$ & $1.34^{+0.26}_{-0.15}$ & 0 (adopted)& $1.8^{+0.06}_{-0.06}$ & $1.7^{+0.05}_{-0.05}$ & $7650^{+160}_{-160}$ & $0.61^{+0.07}_{-0.07}$ & 29 \\
TOI-569 b & CD-41 3255  & 6.556 & $63.8^{+1}_{-1}$ & $0.75^{+0.02}_{-0.02}$ & $< 0.0035 $ & $1.21^{+0.03}_{-0.03}$ & $1.48^{+0.03}_{-0.03}$ & $5705^{+76}_{-76}$ & $0.4^{+0.08}_{-0.08}$ & 30 \\
TOI-1406 b & HD 274870  & 10.574 & $46^{+2.7}_{-2.7}$ & $0.86^{+0.02}_{-0.02}$ & $< 0.039 $ & $1.18^{+0.09}_{-0.09}$ & $1.35^{+0.03}_{-0.03}$ & $6290^{+100}_{-100}$ & $-0.08^{+0.09}_{-0.09}$ & 30 \\
NGTS-7 Ab & NGTS-7  & 0.676 & $75.5^{+3}_{-13.7}$ & $1.38^{+0.13}_{-0.14}$ & 0 (adopted) & $0.48^{+0.13}_{-0.13}$ & $0.61^{+0.06}_{-0.06}$ & $3359^{+106}_{-106}$ & $\cdots$ & 31 \\
2M0535-05 a & V* V2384 Ori  & 9.779 & $56.7^{+4.8}_{-4.8}$ & $6.5^{+0.33}_{-0.33}$ & $0.323^{+0.006}_{-0.006}$ & $\cdots$ & $\cdots$ & $\cdots$ & $\cdots$ & 32 \\
2M0535-05 b & V* V2384 Ori  & 9.779 & $35.6^{+2.8}_{-2.8}$ & $5.00^{+0.25}_{-0.25}$ & $0.323^{+0.006}_{-0.006}$ & $\cdots$ & $\cdots$ & $\cdots$ & $\cdots$ & 32 \\
2M1510 Aa & 2MASS J15104786-2818174   & 20.902 & $40^{+2.9}_{-2.9}$ & $1.53^{+0.15}_{-0.15}$ & $0.309^{+0.022}_{-0.022}$ & $\cdots$ & $\cdots$ & $\cdots$ & $\cdots$ & 33 \\
2M1510 Ab & 2MASS J15104761-2818234  & 20.902 & $39.9^{+2.9}_{-2.9}$ & $1.53^{+0.15}_{-0.15}$ & $0.309^{+0.022}_{-0.022}$ & $\cdots$ & $\cdots$ & $\cdots$ & $\cdots$ & 33 \\
CoRoT-34 b & UCAC4 431-022970  & 2.119 & $71.4^{+8.9}_{-8.6}$ & $1.09^{+0.17}_{-0.16}$ & 0 (adopted) & $1.66^{+0.08}_{-0.15}$ & $1.85^{+0.29}_{-0.25}$ & $7820^{+160}_{-160}$ & $-0.2^{+0.2}_{-0.2}$ & 34 \\
Kepler-503 b & Kepler-503  & 7.258 & $78.6^{+3.1}_{-3.1}$ & $0.96^{+0.06}_{-0.04}$ & $0.025^{+0.014}_{-0.012}$ & $1.154^{+0.047}_{-0.042}$ & $1.764^{+0.08}_{-0.068}$ & $5670^{+100}_{-110}$ & $0.169^{+0.046}_{-0.045}$ & 35 \\
HIP 33609 b & HD 52470  & 39.472 & $68.0^{+7.4}_{-7.4}$ & $1.580^{+0.074}_{-0.070}$ & $0.560^{+0.029}_{-0.031}$ & $2.383^{+0.10}_{-0.095}$ & $1.863^{+0.087}_{-0.082}$ & $10400^{+800}_{-600}$ & $-0.01^{+0.19}_{-0.20}$ & 36 \\
TOI-2336 b & TYC 7317-698-1  & 7.71 & $69.9^{+2.3}_{-2.3}$ & $1.05^{+0.04}_{-0.04}$ & $0.010^{+0.006}_{-0.005}$ & $1.40^{+0.07}_{-0.07}$ & $1.82^{+0.06}_{-0.06}$ & $6433^{+84}_{-84}$ & $0.09^{+0.11}_{-0.11}$ & 37 \\
\enddata
\tablerefs{1: \citet{2022MNRAS.514.4944C}, \citet{2022AJ....163...89C}, 
2: \citet{2023MNRAS.523.6162L},
3: \citet{threebrowndwarfs},
4: \citet{2022AAS...24030521P},
5: \citet{2021MNRAS.505.4956B},
6: \citet{populatingthebrowndwarfandstellarboundary},
7: \citet{2021MNRAS.505.2741A},
8: \citet{2021AJ....162..144A},
9: \citet{2021AJ....161...97C},
10: \citet{2019AJ....157...31Z}
11: \citet{2012ApJ...761..123S}
12: \citet{2010ApJ...718.1353I}
13: \citet{2011ApJ...730...79J}
14: \citet{2018AJ....156..140I}
15: \citet{wasp30b}
16: \citet{2018MNRAS.481.5091H}
17: \citet{corot3b}
18: \citet{corot15b}
19: \citet{corot33b}
20: \citet{kepler39b}
21: \citet{koi189b}
22: \citet{koi205b}
23: \citet{koi415b}
24: \citet{2017AJ....153...15B}
25: \citet{2019AJ....158...38C}
26: \citet{Gillen2017a}
27: \citet{2017AJ....153..131N}
28: \citet{David2019}
29: \citet{2020AJ....159..151S}
30: \citet{2020AJ....160...53C}
31: \citet{2019MNRAS.489.5146J}
32: \citet{2006Natur.440..311S}
33: \citet{2020NatAs...4..650T}
34: \citet{2022MNRAS.516..636S}
35: \citet{2018ApJ...861L...4C}
36: \citet{2023AJ....165..268V}
37: \citet{2023MNRAS.523.6162L}
}
\end{deluxetable}
\end{longrotatetable}

\section{Conclusion}\label{sec:summary}

We characterized three transiting systems: one brown dwarf and two
low-mass stars in the TYC 2010-124-1, UCAC4 515-012898, and BD+45
1593 systems identified by TESS as TOI-2533.01, TOI-5427.01, and
TOI-1712.01.  We showed that the synthesis of Gaia DR3 SB1 solutions
and TESS TOI discoveries offers a promising path to the large-scale,
homogeneous characterization of low-mass stars and massive brown dwarfs.
Gaia DR3 SB1 solutions and TESS transits indicated that the masses
of the transiting objects in all three systems were consistent with
the brown dwarf mass range.  Ground-based precision radial velocity
measurements confirmed that TOI-2533.01 is a transiting brown dwarf with
$M = 72^{+3}_{-3}~M_{\text{Jup}} = 0.069^{+0.003}_{-0.003}~M_\odot$
and that TOI-5427.01 is a transiting very low-mass star with $M =
93^{+2}_{-2}~M_{\text{Jup}} = 0.088^{+0.002}_{-0.002}~M_\odot$.
We validated TOI-1712.01 as a very low-mass star with $M =
82^{+7}_{-7}~M_{\text{Jup}} = 0.079^{+0.007}_{-0.007}~M_\odot$ transiting
a massive subgiant primary in the hierarchical triple system BD+45 1593.
The star BD+45 1593 Ab has a radius nearly 80\% larger than predicted
by theoretical models for an isolated star with a similar masses, solar
composition, and age equal to that of the primary BD+45 1593 Aa.  While
the observation of very-low mass stars with radii larger than predicted 
by theoretical models seems to be ubiquitous and often attributed 
to magnetic fields, radius inflation as large as we observe in BD+45
1593 Ab cannot be attributed to the effect of magnetic fields alone.
We suggest that the intense instellation experienced by BD+45 1593 Ab
heats its photosphere to beyond the photospheric temperature expected for
an isolated star with similar parameters.  This increased photospheric
temperature diminishes the temperature gradient in the outer parts of
BD+45 1593 Ab, suppresses convection, and leads to its inflated radius.
Future studies of systems with Gaia SB1 solutions and Kepler, K2, TESS,
or PLATO transit detections have the potential to provide hundreds of
precise radius inferences for brown dwarfs and very-low mass stars.
These latter data would be critical for the calibration of empirical
mass--radius relations for very low-mass stars that would be very useful
for the calculation of the radii of planets transiting very low-mass
stars like those in the TRAPPIST-1 system.

\section*{Acknowledgments}
We thank the anonymous referee for many very helpful comments.
We thank Zafar Rustamkulov for helpful suggestions about visualizing our 
results. We also thank Daniel Thorngren for helpful insight about our 
findings. This material is based upon work supported by the National Science
Foundation under grant number 2009415 and by the TESS General
Investigator program under NASA grant 80NSSC20K0059.  Samuel N.\
Quinn acknowledges support from the TESS mission via subaward s3449
from MIT.  This work was supported by a NASA WIYN PI Data Award,
administered by the NASA Exoplanet Science Institute.  
This work has made use of data from the European Space Agency (ESA)
mission {\it Gaia} (\url{https://www.cosmos.esa.int/gaia}), processed
by the {\it Gaia} Data Processing and Analysis Consortium (DPAC,
\url{https://www.cosmos.esa.int/web/gaia/dpac/consortium}). Funding for
the DPAC has been provided by national institutions, in particular the
institutions participating in the {\it Gaia} Multilateral Agreement.
This paper includes data collected with the TESS mission, obtained
from the MAST data archive at the Space Telescope Science Institute
(STScI).  Funding for the TESS mission is provided by the NASA Explorer
Program. STScI is operated by the Association of Universities for Research
in Astronomy, Inc., under NASA contract NAS 5–26555.  We acknowledge
the use of public TESS data from pipelines at the TESS Science Office
and at the TESS Science Processing Operations Center.  This research
has made use of the Exoplanet Follow-up Observation Program (ExoFOP;
DOI: 10.26134/ExoFOP5) website, which is operated by the California
Institute of Technology, under contract with the National Aeronautics
and Space Administration under the Exoplanet Exploration Program.
Resources supporting this work were provided by the NASA High-End
Computing (HEC) Program through the NASA Advanced Supercomputing (NAS)
Division at Ames Research Center for the production of the SPOC data
products.  This paper contains data taken with the NEID instrument,
which was funded by the NASA-NSF Exoplanet Observational Research
(NN-EXPLORE) partnership and built by Pennsylvania State University.
NEID is installed on the WIYN telescope, which is operated by the
National Optical Astronomy Observatory, and the NEID archive is operated
by the NASA Exoplanet Science Institute at the California Institute of
Technology.  NN-EXPLORE is managed by the Jet Propulsion Laboratory,
California Institute of Technology under contract with the National
Aeronautics and Space Administration.  Data presented herein were obtained
at the WIYN Observatory from telescope time allocated to NN-EXPLORE
through the scientific partnership of the National  Aeronautics and Space
Administration, the National Science Foundation, and NOIRLab.  The authors
are honored to be permitted to conduct astronomical research on Iolkam
Du’ag (Kitt Peak), a mountain with particular significance to the
Tohono O’odham.  This publication makes use of data products from the
Two Micron All Sky Survey, which is a joint project of the University of
Massachusetts and the Infrared Processing and Analysis Center/California
Institute of Technology, funded by the National Aeronautics and Space
Administration and the National Science Foundation.  This publication
makes use of data products from the Wide-field Infrared Survey Explorer,
which is a joint project of the University of California, Los Angeles,
and the Jet Propulsion Laboratory/California Institute of Technology, and
NEOWISE, which is a project of the Jet Propulsion Laboratory/California
Institute of Technology. WISE and NEOWISE are funded by the National
Aeronautics and Space Administration.  Some/all of the data presented in
this paper were obtained from the Mikulski Archive for Space Telescopes
(MAST). STScI is operated by the Association of Universities for Research
in Astronomy, Inc., under NASA contract NAS5-26555. Support for MAST
for non-HST data is provided by the NASA Office of Space Science via
grant NNX13AC07G and by other grants and contracts.  This research
has made use of the Exoplanet Follow-up Observation Program website,
which is operated by the California Institute of Technology, under
contract with the National Aeronautics and Space Administration under
the Exoplanet Exploration Program.  This research has made use of the
Spanish Virtual Observatory (https://svo.cab.inta-csic.es) project funded
by MCIN/AEI/10.13039/501100011033/ through grant PID2020-112949GB-I00.
This research has made use of the SIMBAD database \citep{Wenger2000},
operated at CDS, Strasbourg, France.  This research made use of
Lightkurve, a Python package for Kepler and TESS data analysis
\citep{2018ascl.soft12013L}.  This work made use of the IPython package
\citep{2007CSE.....9c..21P}.  This research has made use of NASA's
Astrophysics Data System.


\vspace{5mm}
\facilities{ADS, CDS, CTIO:1.5m (CHIRON), CTIO:2MASS, ExoFOP, FLWO:1.5m
(TRES), FLWO:2MASS, Gaia, GALEX, IRSA, MAST, NEOWISE, TESS, WISE, WIYN
(NEID)}

\software{astropy \citep{2013A&A...558A..33A,2018AJ....156..123A,2022ApJ...935..167A},
\texttt{isochrones} \citep{mor15},
\texttt{juliet} \citep{2019MNRAS.490.2262E},
matplotlib \citep{hunter2007matplotlib},
numpy \citep{harris2020array},
pandas \citep{McKinney_2010,reback2020pandas},
PyMultiNest \citep{PyMultiNest},
\texttt{R} \citep{r23},
scipy \citep{2020SciPy-NMeth,jones2001scipy},
\texttt{thejoker} \citep{2017ApJ...837...20P}}

\bibliography{fullbiblio}{}
\bibliographystyle{aasjournal}

\end{CJK*}
\end{document}